\documentclass[12pt,a4paper]{article}
\usepackage{a4wide}
\usepackage{amsmath}
\usepackage{amssymb}
\usepackage{epsfig}
\usepackage{subfigure}
\usepackage{exscale}
\usepackage{float}
\usepackage{bbm}
\usepackage[numbers,sort&compress]{natbib}
\usepackage{pst-plot, pstricks,pst-math}
\usepackage{fancybox,amssymb,color}
\usepackage{graphicx}
\usepackage{pstricks, color, graphicx, epsfig, psfrag}
\usepackage{amsfonts,amsmath,amssymb,slashed}
\usepackage{dsfont}
\usepackage{bbm,bm}
\usepackage{fancyhdr, a4wide}
\usepackage[english]{babel}
\usepackage{subfigure}

\makeatletter
\@addtoreset{equation}{section}
\makeatother

\title{Spontaneous Magnetization of an Ideal Ferromagnet: Beyond Dyson's Analysis}

\author{Christoph P.\ Hofmann$^a$ \\ \\
\normalsize {$^a$ Facultad de Ciencias, Universidad de Colima} \\
\vspace{0.3cm}
\normalsize {Bernal D\'iaz del Castillo 340, Colima C.P.\ 28045, Mexico} \\}

\begin{document}

\maketitle

\begin{abstract} \normalsize

Using the low-energy effective field theory for magnons, we systematically evaluate the partition function
of the O(3) ferromagnet up to three loops. Dyson, in his pioneering microscopic analysis of the Heisenberg
model, showed that the spin-wave interaction starts manifesting itself in the low-temperature expansion of
the spontaneous magnetization of an ideal ferromagnet only at order $T^4$. Although several authors tried to
go beyond Dyson's result, to the best of our knowledge, a fully systematic and rigorous investigation of
higher order terms induced by the spin-wave interaction, has never been achieved. As we demonstrate in the
present paper, it is straightforward to evaluate the partition function of an ideal ferromagnet beyond
Dyson's analysis, using effective Lagrangian techniques. In particular, we show that the next-to-leading
contribution to the spontaneous magnetization resulting from the spin-wave interaction already sets in at
order $T^{9/2}$ -- in contrast to all claims that have appeared before in the literature. Remarkably, the
corresponding coefficient is completely determined by the leading-order effective Lagrangian and is thus
independent of the anisotropies of the cubic lattice. We also consider even higher-order corrections and
thereby solve -- once and for all -- the question of how the spin-wave interaction in an ideal ferromagnet
manifests itself in the spontaneous magnetization beyond the Dyson term.

\end{abstract}
 

\maketitle

\section{Introduction}

\label{intro}

In a landmark paper on the description of ferromagnets at low temperatures \citep{Blo30}, Bloch introduced
the concept of spin waves and identified them as the relevant low-energy degrees of freedom. As an immediate
application, he evaluated the leading coefficient in the low-temperature expansion of the spontaneous
magnetization: this term, corresponding to noninteracting magnons, is of order $T^{3/2}$. As is well-known,
various authors subsequently tried to find the leading term in this series originating from the spin-wave
interaction, ending up with conflicting results: corrections to Bloch's law both of order $T^{7/4}$ and $T^2$
were found \citep{Kra36,Ope37,Sch54,Kra55}. The situation remained rather unclear until Dyson, in his
pioneering analysis of the thermodynamic behavior of an ideal ferromagnet \citep{Dys56}, showed that the
previous results were wrong altogether and that the spin-wave interaction in the spontaneous magnetization
starts manifesting itself only at order $T^4$.

Dyson's motivation was the apparent contradiction between the various results published in the literature.
He successfully solved this paradox by setting up a fairly complicated mathematical machinery -- in his own
words \citep{Dys56}: {\it "The method of the present paper settled the disagreement by showing that both
calculations were wrong"}. In fact, as Dyson states in Ref.~\citep{Dys56}, a third calculation existed that
was also in contradiction with the other two.

Within the last few decades several articles have appeared dealing with the structure of the series for the
spontaneous magnetization {\it beyond} the Dyson term. Various authors, using different methods, have given
their account on what the temperature power of the next-to-leading order term due to the spin-wave
interaction should be and how the general structure of the series beyond Dyson should look like. Not all of
these results that have appeared in the literature over time, however, as we discuss in more detail later
on, are consistent with one another. Our main motivation is thus reminiscent of Dyson's, namely, to
determine which one of these calculations yields the correct low-temperature expansion for the spontaneous
magnetization of an ideal ferromagnet.

Due to its mathematical rigor, Dyson's calculation is not easy to understand and the perturbative scheme
developed for the evaluation of the partition function is fairly complicated. Indeed, after Dyson's
analysis, many authors tried to reproduce and rederive his result with alternative methods in a more
accessible manner \citep{Mor58,Ogu60,KL61,Sza61,Zit65,Wal67,VLP68,CH70,Sza74,YW75,RL79}.
Among these references we would like to point
out the paper by Zittartz \citep{Zit65}, upon which Dyson comments {\it "Zittartz replaced my cumbersome
mathematics by a simple and elegant construction"} \citep{Dys96}. Surprisingly, not all authors were able to
confirm Dyson's result: in particular, a new interaction term in the spontaneous magnetization of order
$T^3$ started to haunt the literature
\citep{Man58,BH60,Eng60,TH62a,SHEB63,Cal63,OH63,Kue69,Tah70,CF73,KG83}
-- later on, this term was recognized as spurious and Dyson's series was confirmed to be correct
\citep{TH62b,Ort64,MT65a,Kef66,Lol87,CH90}.

While all these studies were performed within the framework of microscopic or phenomenological theories
based on the Heisenberg model, in the present work, we will follow another approach which has the virtue of
being completely systematic and model-independent: the method of effective Lagrangians. Within the effective
Lagrangian framework, the structure of the low-temperature expansion of the spontaneous magnetization was
analyzed in Ref.~\citep{Hof02} up to order $T^4$ and Dyson's series was reproduced in a straightforward
manner. In the effective language, as we will see, this corresponds to including Feynman diagrams for the
partition function up to two loops. The effective analysis also readily demonstrated that there is no
interaction term of order $T^3$ in the low-temperature series of the spontaneous magnetization.

In the present work we go beyond Dyson's analysis and explicitly calculate the effect of the spin-wave
interaction beyond $T^4$ in the spontaneous magnetization of an ideal ferromagnet. To the best of our
knowledge, this is the first time that the structure of this power series is given in a fully systematic and
rigorous way. Going beyond Dyson's analysis then means that, in the effective Lagrangian framework, we have
to consider Feynman diagrams up to three-loop order in the perturbative expansion of the partition function.
As it turns out, in the spontaneous magnetization of an ideal ferromagnet, the next-to-leading interaction
term already sets in at order $T^{9/2}$ -- remarkably, the corresponding coefficient is completely
determined by the two low-energy coupling constants of the leading-order effective Lagrangian
${\cal L}^2_{eff}$. It does not involve any higher-order effective constants from ${\cal L}^4_{eff}$ where
the anisotropies of the cubic lattice start showing up.

Although several authors have also discussed the structure of temperature powers beyond the $T^4$-term,
their conclusions are in contradiction with the systematic effective field-theory approach and therefore
erroneous. In particular, to the best of our knowledge, none of the existing calculations ended up with an
{\it interaction} term of order $T^{9/2}$, which in fact represents the leading correction to Dyson's result.

Within the effective Lagrangian framework, we then analyze the general structure of even higher-order
corrections in the spontaneous magnetization originating from the spin-wave interaction, and show that these
are of order $T^5, T^{11/2}, T^6, \dots$ -- again contradicting earlier calculations that have appeared in
the literature.

The effective Lagrangian method is based on an analysis of the symmetry properties of the underlying theory,
i.e., the Heisenberg model in our case, and can universally be applied to systems with a spontaneously
broken symmetry. It is formulated in terms of Goldstone boson fields which represent the dominant low-energy
degrees of freedom. The effective Lagrangian method is very well established in particle physics, where the
low-energy effective theory for quantum chromodynamics -- chiral perturbation theory -- has been constructed
a long time ago \citep{Wei79,GL85}. There, we are dealing with a spontaneously broken chiral symmetry and the
corresponding Goldstone bosons are the pseudoscalar mesons. Spontaneous symmetry breaking is also a common
phenomenon in condensed matter physics and the effective Lagrangian method has in fact been transferred to
this domain in Ref.~\citep{Leu94a}: Magnons and phonons, e.g., are the Goldstone bosons resulting from a
spontaneously broken spin rotation symmetry $O(3) \to O(2)$ and a spontaneously broken translation symmetry,
respectively. In particular, the leading-order effective Lagrangian for the O(3) ferromagnet was constructed
in Ref.~\citep{Leu94a} and the extension to higher orders in the derivative expansion was performed in
Refs.~\citep{RS99a,Hof02}. 

The paper is organized as follows. Since the systematic effective Lagrangian method is still not very well
known within the condensed matter community, in Sec.~\ref{method} we give a brief outline of the method,
having in mind the ferromagnet as specific system. In Sec.~\ref{detailsA} we briefly review the evaluation
of the partition function of an ideal ferromagnet up to order $T^5$. We then go beyond Dyson's analysis and
extend the evaluation to order $T^{11/2}$ in Sec.~\ref{detailsB}. While the renormalization up to order $T^5$
is straightforward, the handling of ultraviolet divergences at order $T^{11/2}$ is more involved and is
considered in detail in Sec.~\ref{renorm}. The low-temperature expansion of the partition function and
various thermodynamic quantities is given in Sec.~\ref{thermodynamics}. Our main result -- the
low-temperature series for the spontaneous magnetization of an ideal ferromagnet beyond Dyson's analysis --
is presented in Sec.~\ref{spontaneousMagnetization}. Here we also compare our results with the condensed
matter literature. While our conclusions are presented in Sec.~\ref{conclusions}, details on the numerical
evaluation of a specific three-loop graph are discussed in Appendix \ref{appendix A}.

We would like to provide the interested reader with a list of publications that deal with applications of
the effective Lagrangian method to condensed matter systems. Applications to systems exhibiting collective
magnetic behavior include spin-wave scattering processes \citep{Hof99a}, spin-wave mediated nonreciprocal
effects in antiferromagnets \citep{RS99b}, antiferromagnets at finite volume \citep{HL90,HN91,HN93,Bie93}
and finite temperature \citep{Hof99b,Hof10}, spin waves in canted phases \citep{RS00} and antiferromagnets
in two dimensions doped with charge carriers \citep{KMW05,BKMPW06,BKPW06,BHKPW07,BHKMPW07,JKHW09}. Further
applications include phonons \citep{Leu97}, SO(5) invariance and high-$T_c$-superconductivity \citep{BL98}
as well as supersolids \citep{Son94}. Pedagogic introductions to the effective Lagrangian method may be
found in Refs.~\citep{Brau10,Bur07,Goi04,Sch03,Man97,Leu95,Eck95}.

In particular, we would like to point out that in a recent article on an analytically solvable microscopic
model for a hole-doped ferromagnet in 1+1 dimensions \citep{GHKW10}, the correctness of the effective field
theory approach was demonstrated by comparing the effective theory predictions with the microscopic
calculation. Likewise, in a series of high-accuracy investigations of the antiferromagnetic
spin-$\frac{1}{2}$ quantum Heisenberg model on a square lattice using the loop-cluster algorithm
\citep{WJ94,GHJNW09,JW11,GHJPSW11}, the Monte Carlo data were confronted with the analytic predictions of
magnon chiral perturbation theory and the low-energy constants were extracted with permille accuracy. All
these tests unambiguously demonstrate that the effective Lagrangian approach provides a rigorous and
systematic derivative expansion for both ferromagnetic an antiferromagnetic systems.

\section{Systematic Low-Energy Effective Field Theory for Ferromagnetic Magnons}
\label{method}

The effective Lagrangian method is based on a symmetry analysis of the underlying system. In the present
case we study ferromagnets, which are described by the Heisenberg model. Nevertheless, the effective
field-theory predictions are model-independent and universal, as they are valid for any system displaying
the same symmetries as the Heisenberg ferromagnet. Microscopic details of the system are taken into account
through a few low-energy coupling constants in the effective Lagrangian. Symmetry does not fix the actual
numerical values of these couplings -- in general, these have to be determined experimentally or in a
numerical simulation of the underlying model. Symmetry, however, does unambiguously determine the derivative
structure of the terms in the effective Lagrangian.

The most important symmetry in the present case is the spontaneously broken spin rotation symmetry: Whereas
the Heisenberg model,
\begin{equation}
\label{HeisenbergModel}
{\cal H}_0 = -J \sum_{n.n.} {\vec S}_m \cdot {\vec S}_n \, , \qquad J=const. \, , 
\end{equation}
is invariant under global O(3) spin rotations, the ground state of the ferromagnet ($J>0$) is invariant
under the subgroup O(2) only. According to the nonrelativistic Goldstone  theorem
\citep{Lan66,GHK68,CN76,Sch01,Bra07}, we then have one type of spin-wave excitation -- or one magnon
particle -- in the low-energy spectrum of the ferromagnet which obeys a quadratic dispersion relation.

The interaction between an external constant magnetic field ${\vec H}=(0,0,H), H>0$ and the spin degrees of
freedom is taken into account through the Zeeman term. In the corresponding extension of the Heisenberg
model,
\begin{equation}
{\cal H} = {\cal H}_0 - \mu \sum_n {\vec S}_n \cdot {\vec H} \, ,
\end{equation}
the magnetic field couples to the vector of the total spin. The above Hamiltonian, defined on a cubic
lattice with purely isotropic exchange coupling between nearest neighbors, represents what Dyson called
{\it ideal} ferromagnet. 

Apart from internal symmetries we also have to consider the various space-time symmetries. Compared to
particle physics where we have Lorentz invariance, the situation is more complicated in condensed matter
physics, because the center of mass system represents a preferred frame of reference. Moreover, the crystal
lattice singles out preferred directions, such that the effective Lagrangian need not be rotationally
invariant. In the case of cubic geometry, however, it has been shown that the anisotropies of the lattice
start manifesting themselves at higher orders of the derivative expansion \citep{HN93} -- the leading-order
effective Lagrangian is thus invariant under space rotations. Moreover, as the effective analysis refers to
large wavelengths, it does not resolve the microscopic structure of the crystal: the system appears
homogeneous and the effective Lagrangian is also invariant under translations.

The idea underlying the construction of effective Lagrangians is straightforward \citep{Leu94b}: One writes
down the most general expression consistent with the space-time symmetries and the internal, spontaneously
broken symmetry G of the underlying system in terms of Goldstone boson fields $U^a(x), a = 1, \dots$,
dim(G)-dim(H), where the group H refers to the symmetry group of the ground state. The effective Lagrangian
then consists of a string of terms involving an increasing number of derivatives or, equivalently, amounts
to an expansion in powers of the momentum. Furthermore, the effective Lagrangian method allows to
systematically take into account interactions which explicitly break the symmetry G of the underlying model,
provided that they can be treated as perturbations. In the present case we will include a weak external
magnetic field ${\vec H}$.

For the O(3) ferromagnet, the leading-order effective Lagrangian is of order $p^2$ and takes the form
\citep{Leu94a}
\label{leadingLagrangian}
\begin{equation}
{\cal L}^2_{eff} = \Sigma \frac{\epsilon_{ab} {\partial}_0 U^a U^b}{1+ U^3}
+ \Sigma \mu H U^3 - \frac{1}{2} F^2 {\partial}_r U^i {\partial}_r U^i \, .
\end{equation}
The two real components of the magnon field, $U^a (a=1,2)$ are the first two components of the
three-dimensional unit vector $U^i = (U^a, U^3)$, which transforms with the vector representation of the
rotation group. While the structure of the above terms is unambiguously determined by the symmetries of the
underlying theory, at this order, we have two a priori unknown low-energy constants: the spontaneous
magnetization $\Sigma$ and the constant $F$. The above Lagrangian leads to a quadratic dispersion relation
\begin{equation}
\omega({\vec k}) = \gamma {\vec k}^2 + {\cal O}( { |{\vec k}| }^4) \, , \quad
\gamma \equiv \frac{F^2}{\Sigma} \, ,
\end{equation}
obeyed by ferromagnetic magnons. It is important to note that one temporal derivative (${\partial}_0$) is on
the same footing as two spatial derivatives (${\partial}_r {\partial}_r$)  -- in the derivative expansion,
two powers of momentum thus count as only one power of energy or temperature: $k^2 \propto \omega, T$.

Dyson evaluated the low-temperature expansion of the spontaneous magnetization up to terms of order $T^4$
or, equivalently, the partition function up to order $T^5$. This then means that, in the effective
Lagrangian framework, we have to consider the expansion of the partition function up to order $p^{10}$. This
calculation was performed in Ref.~\citep{Hof02}. In the present work, we go one step further and consider
the expansion beyond Dyson's analysis, taking into account diagrams of order $p^{11}$. As it turns out, the
corresponding contributions lead to a spin-wave interaction term of order $T^{9/2}$ in the spontaneous
magnetization.

The effective Lagrangian method provides us with a simultaneous expansion of physical quantities in powers
of the momenta and of the external fields. The essential point is that, to a given order in the low-energy
expansion, only a finite number of effective coupling constants and only a finite number of graphs
contribute. The leading terms stem from tree graphs, whereas loop graphs only manifest themselves at higher
orders in the derivative expansion \citep{Wei79}. So the question arises as to what order in the effective
expansion we have to go -- i.e., how many derivatives in the effective Lagrangian we have to include and how
many loops we have to consider -- if we want to evaluate the partition function of a ferromagnet up to order
$p^{11}$. 

\begin{figure}[t]
\begin{center}
\epsfig{file=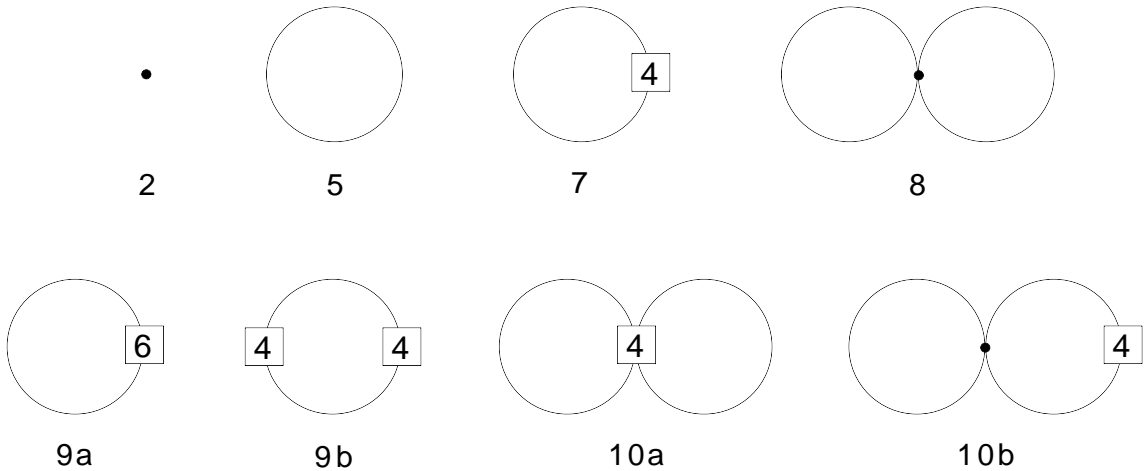,width=15cm}
\end{center}
{\caption{Feynman graphs related to the low-temperature expansion of the partition function for a
ferromagnet up to order $p^{10}$ in dimension $d$=3+1. The numbers attached to the vertices refer to the
piece of the effective Lagrangian they come from. Vertices associated with the leading term
${\cal L}^2_{eff}$ are denoted by a dot. Note that ferromagnetic loops are suppressed by three momentum
powers in $d$=3+1.}
\label{figure1} }
\end{figure}

\begin{figure}[t]
\begin{center}
\epsfig{file=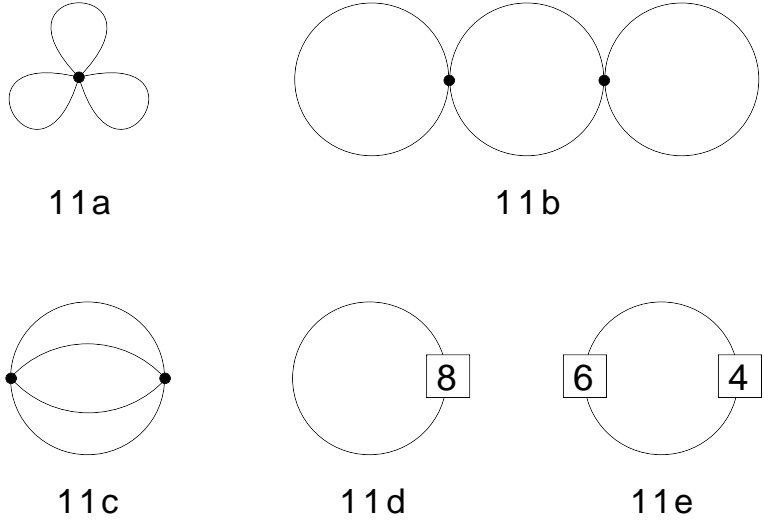,width=10cm}
\end{center}
{\caption{Feynman graphs related to the low-temperature expansion of the partition function for a
ferromagnet at order $p^{11}$ in dimension $d$=3+1. The numbers attached to the vertices refer to the piece
of the effective Lagrangian they come from. Vertices associated with the leading term ${\cal L}^2_{eff}$ are
denoted by a dot.}
\label{figure2} }
\end{figure}

While loops are suppressed by {\it two} momentum powers in a Lorentz-invariant framework, it was shown in
Ref.~\citep{Hof02} that loop corrections involving ferromagnetic magnons are suppressed by {\it three}
momentum powers \footnote{We are considering the case of four space-time dimensions. If one lowers the
spatial dimension, loops are less suppressed: Loops for ferromagnetic magnons in d=2+1, e.g., are suppressed
by two momentum powers, while loops for antiferromagnetic magnons in d=2+1 are suppressed by one momentum
power only.}. Up to order $p^{10}$, as depicted in Fig.~\ref{figure1}, we thus have to consider graphs which
involve two loops at most and have to take into account pieces of the effective Lagrangian involving up to
six derivatives. At order $p^{11}$, as depicted in Fig.~\ref{figure2}, three-loop graphs start to show up.
At the same time we also have two one-loop graphs which involve vertices from higher-order pieces of the
effective Lagrangian: Diagram 11d contains a vertex from ${\cal L}^8_{eff}$, while diagram 11e contains
insertions from ${\cal L}^4_{eff}$ and ${\cal L}^6_{eff}$. These five graphs represent the additional
diagrams we have to evaluate when we go one step beyond Dyson's analysis.

We now address the question regarding the explicit structure of the pieces
${\cal L}^4_{eff}$, ${\cal L}^6_{eff}$ and ${\cal L}^8_{eff}$. First of all, note that there are no
contributions to the effective Lagrangian leading to odd momentum powers: The pieces
${\cal L}^3_{eff}, {\cal L}^5_{eff}, \dots$ necessarily involve terms with an odd number of space derivatives
like
\begin{equation}
c_{abc} \, \epsilon_{r st} \, {\partial}_r U^a \, {\partial}_s U^b \, {\partial}_t U^c \, ,
\end{equation}
which are excluded by parity -- parity is a discrete symmetry of the underlying Heisenberg model that has to
be respected by the effective Lagrangian.

The next-to-leading order Lagrangian is thus of order $p^4$. It contains terms with two time derivatives,
terms with one time and two space derivatives, and terms with four space derivatives. The time derivatives
along with the magnetic field, however, can be eliminated with the equation of motion, such that
${\cal L}^4_{eff}$ takes the form \citep{Hof02}
\begin{equation}
\label{Leff4}
{\cal L}^4_{eff} = l_1 {( {\partial}_r U^i {\partial}_r U^i )}^2 +
l_2 {( {\partial}_r U^i {\partial}_s U^i )}^2 + l_3 \Delta U^i \Delta U^i \, ,
\end{equation}
where $\Delta$ denotes the Laplace operator in three dimensions. The next-to-leading order effective
Lagrangian hence involves the three effective coupling constants $l_1, l_2$ and $l_3$.

An inspection of the diagrams in Figs.~\ref{figure1} and \ref{figure2} reveals that insertions from
${\cal L}^6_{eff}$ and ${\cal L}^8_{eff}$ only appear in one-loop graphs: the only terms we need are thus
quadratic in the magnon field. Eliminating again time derivatives and terms involving the magnetic field,
the pieces relevant for our calculation are
\begin{equation}
\label{Leff68}
{\cal L}^6_{eff} = c_1 U^i {\Delta}^3 U^i \, , \qquad {\cal L}^8_{eff}
= d_1 U^i {\Delta}^4 U^i \, .
\end{equation}

We conclude this section with a remark concerning effects induced by the anisotropy of the lattice.
Regarding the cubic lattice, we have mentioned that the anisotropies start manifesting themselves at the
four-derivative level: the pieces ${\cal L}^4_{eff}$, ${\cal L}^6_{eff}$ and ${\cal L}^8_{eff}$ indeed contain
additional terms -- not displayed in Eqs.(\ref{Leff4}) and (\ref{Leff68}) -- which are not invariant under
space rotations, but still invariant under the discrete rotation and reflexion symmetries of the cubic
lattice, such as
\begin{equation}
\label{extraTerm1}
\sum_{s=1,2,3} \, {\partial}_s {\partial}_s U^i \, {\partial}_s {\partial}_s
U^i \, .
\end{equation}
In the present analysis, however, we neglect these extra terms and assume space rotation invariance up to
order $p^8$. The conclusions of the present paper regarding the manifestation of the spin-wave interaction
in the partition function are not affected by this idealization: According to Fig.~\ref{figure2}, the
interaction contribution beyond Dyson is determined by the three-loop graphs of order $p^{11}$: These
graphs only involve the leading-order Lagrangian ${\cal L}^2_{eff}$ which is perfectly invariant under space
rotations.

\section{Evaluation of the Partition Function}
\label{eval}

The low-temperature expansion of the partition function for the O(3) ferromagnet was evaluated in
Ref.~\citep{Hof02} up to order $p^{10}$. In Sec.~\ref{detailsA} we briefly review some essential features of
that calculation. In Sec.~\ref{detailsB} we then extend the evaluation of the partition function to order
$p^{11}$. For a review of the effective Lagrangian method at non-zero temperature, the interested reader may
consult Ref.~\citep{Leu89}. For a general review of field theory at finite temperature, see
Refs.~\citep{LW87,Kap98,Smi97}.

\subsection{Evaluation up to order $\mathbf{p^{10}}$}
\label{detailsA}

In finite-temperature field theory the partition function is represented as a Euclidean functional integral
\begin{equation}
\label{TempExp}
\mbox{Tr} \, [\exp(- {\cal H}/T)] \, = \, \int [dU] \, \exp \Big(- {\int}_{\!
\! \! {\cal T}} \! \! d^4x \, {\cal L}_{eff}\Big) \, .
\end{equation}
The integration is performed over all field configurations which are periodic in the Euclidean time
direction, $U({\vec x}, x_4 + \beta) = U({\vec x}, x_4)$ with $\beta \equiv 1/T$. The periodicity condition
imposed on the magnon fields also reflects itself in the thermal propagator
\begin{equation}
\label{ThermalPropagator}
G(x) \, = \, \sum_{n \,= \, - \infty}^{\infty} \Delta({\vec x}, x_4 + n \beta) \, ,
\end{equation}
where $\Delta(x)$ is the Euclidean propagator at zero temperature,
\begin{equation}
\label{Propagator}
\Delta (x) \, = \, \int \! \, \frac{d k_4 d^3\!k}{(2\pi)^4}
\frac{e^{i{\vec k}{\vec x} - i k_4 x_4}}{\gamma {\vec k}^2 - i k_4 + \mu H}
= \Theta(x_4) \int \frac{d^3\!k}{(2\pi)^3} \, 
e^{i{\vec k}{\vec x} - \gamma {\vec k}^2 x_4 - \mu H x_4} \, . 
\end{equation}
An explicit representation for the thermal propagator, dimensionally regularized in the spatial dimension
$d_s$, is
\begin{equation}
\label{ThermProp}
G(x) \ = \ \frac{1}{(2{\pi})^{d_s}} \, \Big(\frac{{\pi}}{\gamma}\Big)^{\frac{d_s}{2}}
\sum^{\infty}_{n \, = \, - \infty} \frac{1}{x_n^{\frac{d_s}{2}}} \,
\exp{\Big(- \frac{{\vec x}^2}{4 \gamma x_n} \, - \, \mu H x_n \Big)} \,
\Theta (x_n) \, ,
\end{equation}
with
\begin{equation}
x_n \, \equiv \, x_4 + n \beta \, .
\end{equation}
We restrict ourselves to the infinite volume limit and evaluate the free energy density $z$, defined by
\begin{equation}
\label{freeEnergyDensity}
z \ = \ - \, T \, \lim_{L\to\infty} L^{-3} \, \ln \, [\mbox{Tr}
\exp(-{\cal H}/T)] \, .
\end{equation}

In the evaluation of the various Feynman diagrams, we will repeatedly be dealing with thermal propagators
(and space derivatives thereof), which have to be evaluated at the origin. It is convenient to introduce the
following notation,
\begin{equation}
\label{definitionsThermProp}
G_1 \equiv \Big[ G(x) \Big]_{x=0} \, , \quad
G_{\Delta} \equiv \Big[ {\Delta} G(x) \Big]_{x=0} \, , \quad
G_{\Delta^n} \equiv \Big[ {\Delta}^n G(x) \Big]_{x=0} \, ,
\end{equation}
where $\Delta$ represents the Laplace operator in the spatial dimensions -- no confusion should occur with
$\Delta (x)$, which denotes the zero-temperature propagator.

The quantities $G_1$, $G_{\Delta}$, as well as thermal propagators involving higher-order space derivatives,
are split into a finite piece, which is temperature dependent, and a divergent piece, which is temperature
independent,
\begin{equation}
\label{DecompositionPropagator}
G_1 \; = \; G^T_1 \, + \, G^0_1 \, , \qquad G_{\Delta} \; = \; G^T_{\Delta} \,
+ \, G^0_{\Delta} \, .
\end{equation}
The explicit expressions can be found in Ref.~\citep{Hof02} and will not be given here. Rather, we would
like to point out two important observations, which lead to a substantial simplification of the
renormalization procedure. First, the temperature-independent pieces $G^0_1, G^0_{\Delta}, \dots$ are all
related to momentum integrals of the form
\begin{equation}
\label{RuleDimReg}
\int \! \, d^{d_s}\!k \, \Big( {\vec k}^2 \Big)^m
\exp\Big[ - \gamma x_4 {\vec k}^2 - x_4 \mu H \Big] \, , \qquad m = 0, 1, 2,
\ldots \, ,
\end{equation}
which are proportional to
\begin{equation}
\label{RuleDimReg2}
\frac{\exp[- x_4 \mu H]}{(\gamma x_4)^{m+\frac{d_s}{2}}} \, \,
\Gamma\Big(m+\frac{d_s}{2}\Big) \, .
\end{equation}
In dimensional regularization these expressions vanish altogether: $G^0_1, G^0_{\Delta}$, and
zero-temperature propagators involving higher-order space derivatives do not contribute in the limit
$d_s \! \to \! 3$.

The second observation concerns the fact that, up to order $p^{10}$, the individual contributions to the
free energy density from the various diagrams factorize into products of thermal propagators (involving
space derivatives or derivatives with respect to the magnetic field), which all have to be evaluated at the
origin. As an example consider the two-loop graph 10a which yields the contribution
\begin{equation}
\label{z(10a)}
z_{10a} = - \frac{2}{3 {\Sigma}^2} (8 l_1 + 6 l_2 + 5 l_3) \, G_{\Delta}
G_{\Delta} \, - \frac{2 \, l_3}{{\Sigma}^2} \, G_1 \, G_{{\Delta}^2}.
\end{equation}
According to the first observation regarding dimensional regularization, it is then clear that in the two
products of thermal propagators above only the fully temperature-dependent pieces --
$G^T_{\Delta} G^T_{\Delta}$ and $G^T_1 G^T_{{\Delta}^2}$ -- are nonzero, whereas any other terms involving
temperature-independent pieces of propagators vanish identically. We thus conclude that, using dimensional
regularization, the renormalization of the partition function up to order $p^{10}$ is quite trivial. As we
will see in the next subsection, the renormalization at the three-loop level, on the other hand, is more
complicated, but still perfectly feasible within the effective field theory framework.

Without going into more details (the interested reader may consult Ref.~\citep{Hof02}), we present the final
result for the free energy density of the O(3) ferromagnet up to order $p^{10}$:
\begin{eqnarray}
\label{FreeCollect}
z = & - & \Sigma \mu H \; - \; \frac{1}{8 {\pi}^{\frac{3}{2}}
{\gamma}^{\frac{3}{2}}} \
T^{\frac{5}{2}} \sum^{\infty}_{n=1} \frac{e^{- \mu H n \beta}}{n^{\frac{5}{2}}}
\; - \; \frac{15 \,l_3}{16 {\pi}^{\frac{3}{2}} \Sigma {\gamma}^{\frac{7}{2}}} \
T^{\frac{7}{2}} \sum^{\infty}_{n=1} \frac{e^{- \mu H n \beta}}{n^{\frac{7}{2}}}
\nonumber \\
& - & \frac{105}{32 {\pi}^{\frac{3}{2}} \Sigma {\gamma}^{\frac{9}{2}}} \,
\Bigg( \frac{9 l^2_3}{2 \gamma \Sigma} - c_1 \Bigg) \ T^{\frac{9}{2}}
\sum^{\infty}_{n=1} \frac{e^{- \mu H n \beta}}{n^{\frac{9}{2}}}
\nonumber \\
& - & \frac{3 (8 l_1 + 6 l_2 + 5 l_3)}{128 {\pi}^3 {\Sigma}^2 {\gamma}^5} \ T^5
{\Bigg\{ \sum^{\infty}_{n=1} \frac{e^{- \mu H n\beta}}{n^{\frac{5}{2}}} \Bigg\}}^2
+ {\cal O}(T^{\frac{11}{2}}) \, .
\end{eqnarray}
The first term in this series does not depend on temperature and originates from the tree graph 2 (see
Fig.~\ref{figure1}). The terms which involve half-integer powers of the temperature -- $T^{5/2}, T^{7/2}$ and
$T^{9/2}$, respectively -- arise from the one-loop graphs displayed in Fig.~\ref{figure1}. They all
contribute to the free energy density of noninteracting magnons. Remarkably, up to order $p^{10}$, there is
only one term in the above series -- the contribution of order $T^5$ coming from the two-loop graphs 10a and
10b -- which is due to the magnon-magnon interaction.

In particular, there is no term of order $T^4$ in the above series for the free energy density: The two-loop
graph 8, which would be the only candidate to yield such a contribution, is proportional to single space
derivatives of the thermal propagator evaluated at the origin:
\begin{equation}
\label{z(8)}
z_8 \propto {[\partial_r G(x)]}_{x=0} \,  {[\partial_r G(x)]}_{x=0} = 0 \, .
\end{equation}
This contribution vanishes due to space rotation invariance of the leading-order effective Lagrangian.

\subsection{Evaluation at order $\mathbf{p^{11}}$}
\label{detailsB}

According to Fig.~\ref{figure2} we have a total of five diagrams at order $p^{11}$. We first consider the
two one-loop graphs which involve vertices from ${\cal L}^4_{eff}$, ${\cal L}^6_{eff}$ and ${\cal L}^8_{eff}$.
For graph 11d we obtain
\begin{equation}
\label{z(11d)}
z_{11d} \ = \ - \frac{2 \, d_1}{\Sigma} \, G_{{\Delta}^4} \, ,
\end{equation}
yielding the temperature-dependent contribution
\begin{equation}
\label{z(11d)T}
z^T_{11d} \ = \ - \frac{945 \, d_1}{64 {\pi}^{\frac{3}{2}} \Sigma
{\gamma}^{\frac{11}{2}}} \, T^{\frac{11}{2}} \sum^{\infty}_{n=1}
\frac{e^{- \mu H n \beta}}{n^{\frac{11}{2}}} \, \, .
\end{equation}
Graph 11e is proportional to an integral over the torus ${\cal T} = {\cal R}^{d_s} \times S^1$, with circle
$S^1$ defined by $- \beta / 2 \leq x_4 \leq \beta / 2$, and involves a product of two thermal propagators,
\begin{equation}
\label{z(11e)}
z_{11e} \, = \, - \frac{4 \, l_3 c_1}{{\Sigma}^2} \, {\int}_{\! \! \! {\cal T}}
\! \! d^{d_s+1}x \, {\Delta}^2 G(x) \, {\Delta}^3 G(-x) \, .
\end{equation}
This integral, however, can be reduced to an expression involving one propagator only, using the relation
\citep{Hof02}
\begin{equation}
\label{2->1Delta4}
\Bigg[ {\Delta}^{(m+n)} \frac{\partial G(x)}{\partial (\mu H)} \Bigg]_{x=0} \,
= \, - {\int}_{\! \! \! {\cal T}} \! \! d^{d_s+1}y \, {\Delta}^m G(-y) \,
{\Delta}^n G(y) \, .
\end{equation}
We then end up with
\begin{equation}
z_{11e} \, = \, \frac{4 \, l_3 c_1}{{\Sigma}^2} \, \Bigg[ {\Delta}^5
\frac{\partial G(x)}{\partial (\mu H)} \Bigg]_{x=0} \, .
\end{equation}
Accordingly, the temperature-dependent part of graph 11e reads
\begin{equation}
\label{z(11e)T}
z^T_{11e} \ = \  \frac{10395 \, l_3 c_1}{64 {\pi}^{\frac{3}{2}} {\Sigma}^2
{\gamma}^{\frac{13}{2}}} \, T^{\frac{11}{2}} \sum^{\infty}_{n=1}
\frac{e^{- \mu H n \beta}}{n^{\frac{11}{2}}} \, \, .
\end{equation}

We now turn to the three-loop graphs -- note that they exclusively contain vertices from the leading-order
Lagrangian ${\cal L}^2_{eff}$. Graph 11a factorizes into a product of three thermal propagators (and space
derivatives thereof), to be evaluated at the origin,
\begin{equation}
z_{11a} \, = \, - \frac{F^2}{{\Sigma}^3} \, G_{\Delta} {(G_1)}^2 \, .
\end{equation}
The subsequent three-loop graph 11b, remarkably, does not contribute to the partition function,
\begin{equation}
\label{z(8)}
z_{11b} = 0 \, .
\end{equation}
As it was the case for the two-loop graph 8, the three-loop graph 11b is identically zero.

Finally, for the cateye graph 11c we get
\begin{equation}
\label{cateye}
z_{11c} \, = \, - \frac{F^4}{2{\Sigma}^4} \, J \, + \, \frac{F^2}{{\Sigma}^3 }
G_{\Delta} {(G_1)}^2 \, .
\end{equation}
The expression $J$ stands for the following integral over the torus involving a product of four thermal
propagators
\begin{equation}
\label{integralJ}
J \, = {\int}_{\! \! \! {\cal T}} \! \! d^{d_s+1}x \, {\partial}_r G \,
{\partial}_r G \, {\partial}_s {\tilde G} \, {\partial}_s {\tilde G} \, ,
\end{equation}
where we have used the notation
\begin{equation}
G = G(x) \, , \qquad {\tilde G} = G(-x) \, .
\end{equation}
Note that the second term in (\ref{cateye}) cancels the contribution from graph 11a, such that the overall
contribution from the three-loop graphs is the one proportional to the integral $J$. Remarkably, unlike all
other pieces in the free energy density up to order $p^{11}$, this quantity is not just a product of thermal
propagators (or derivatives thereof) to be evaluated at the origin. The remaining task will be the
renormalization and the numerical evaluation of this integral which contains a total of four infinite sums.
In the next section and in Appendix \ref{appendix A} we address this problem in detail.

Leaving aside these technical issues for a moment, we note that the cateye graph of order $p^{11}$ will lead
to a term of order $T^{11/2}$ in the free energy density,
\begin{equation}
J \propto T^{\frac{11}{2}} \, .
\end{equation}
Hence the spin-wave interaction in the low-temperature series of the free energy density -- beyond Dyson's
$T^5$-term -- already manifests itself at order $T^{11/2}$. It is remarkable that this contribution is
exclusively determined by the symmetries of the leading-order effective Lagrangian ${\cal L}^2_{eff}$ which
involves the two couplings $\Sigma$ and $F$ -- the spin-wave interaction at this order is not affected by
the anisotropies of the cubic lattice.

\section{Renormalization of the Cateye Graph}
\label{renorm}

Using dimensional regularization it was straightforward to extract the finite pieces in the partition
function up to two-loop order $p^{10}$. The renormalization of the three-loop graph 11c, on the other hand,
is more involved. We will follow the procedure outlined in Ref.~\citep{GL89}, where the same graph was
considered in the context of a Lorentz-invariant effective field theory.

To analyze the integral
\begin{displaymath}
J \, = {\int}_{\! \! \! {\cal T}} \! \! d^{d_s+1}x \, {\partial}_r G \,
{\partial}_r G \, {\partial}_s {\tilde G} \, {\partial}_s {\tilde G}
\end{displaymath}
in the limit $d_s \! \to \! 3$, we split the thermal propagator into two pieces
\begin{equation}
G(x) = G^T(x) + \Delta(x) \, .
\end{equation}
The ultraviolet singularities are contained in the zero-temperature propagator $\Delta(x)$, whereas the
temperature-dependent part $G^T(x)$ is finite as $d_s \! \to \! 3$. Note that, if we restrict ourselves to
the origin, we reproduce the first relation of Eq.(\ref{DecompositionPropagator}).

Inserting the above decomposition into the integral $J$, we end up with nine terms that can be grouped into
the following six classes:\footnote{For simplicity we do not display the derivatives.}
\begin{eqnarray}
\label{SixTypes}
A:& & G^T(x) \, G^T(x) \, G^T(-x) \, G^T(-x) ,
\nonumber \\
B:& & \Delta(x) \, G^T(x) \, G^T(-x) \, G^T(-x) \, ,
\quad G^T(x) \, G^T(x) \, \Delta(-x) \, G^T(-x) , \nonumber \\
C:& & {\Delta}^2(x) \, G^T(-x) \, G^T(-x) \, , \quad
G^T(x) \, G^T(x)  \,\Delta^2(-x), \nonumber \\
D:& & \Delta(x) \, G^T(x) \, \Delta(-x) \, G^T(-x)\, ,
\nonumber \\
E:& & {\Delta}^2(x) \, \Delta(-x) \, G^T(-x) \, ,
\quad \Delta(x) \, G^T(x) \, {\Delta}^2(-x) \, , \nonumber \\
F:& & {\Delta}^2(x) \, {\Delta}^2(-x).
\end{eqnarray}
Terms of the classes $D, E$ and $F$ vanish identically since the product $\Delta(x) \Delta(-x)$ of 
zero-temperature propagators involves the combination $\Theta(x_4) \Theta(-x_4)$. The maximum number of
$\Theta$-functions a given term can contain -- in order not to be zero -- is two. Moreover, the arguments of
the two $\Theta$-functions have to coincide as it is the case with the terms of class $C$. We thus have to
consider the cases $A, B$ and $C$. 

The integral over the torus involving contributions of classes $A$ and $B$,
\begin{equation}
{\int}_{\! \! \! {\cal T}} \! \! d^{d_s+1}x \, \Big( {\partial}_{r} G^T {\partial}_{r}
G^T {\partial}_{s} {\tilde G}^T {\partial}_{s} {\tilde G}^T
+ 4 \, {\partial}_{r} \Delta \, {\partial}_{r} G^T {\partial}_{s} {\tilde G}^T
 {\partial}_{s} {\tilde G}^T \Big) \, ,
\end{equation}
converges at $d_s=3$.

Terms of class $C$, however, do lead to an ultraviolet-divergent integral. Consider, e.g., the term
\begin{equation}
\label{termClassC}
{\partial}_r \Delta(x) \, {\partial}_r \Delta(x) \, {\partial}_s G^T(-x) \, {\partial}_s G^T(-x) \, ,
\end{equation}
where we now have displayed the derivatives. For the zero-temperature piece ${\partial}_r \Delta(x)$ we have
\begin{equation}
{\partial}_r \Delta(x) \propto \frac{x^r}{{x_4}^{\frac{5}{2}}} \,
\exp\Big[ - \frac{{\vec x}^2}{4 \gamma x_4} \Big] \, .
\end{equation}
The Taylor series of the function ${\partial}_s G^T(-x)$, evaluated at the origin, starts with a term linear
in ${\vec x}$,
\begin{equation}
\label{Taylor}
{\partial}_s G^T(-x) \, = \, {\partial}_{\alpha s} G^T(-x)|_{x=0} \, x^{\alpha} + {\cal O}({\vec x}^3) \, .
\end{equation}
Inserting this term into Eq.(\ref{termClassC}), we end up with the following contribution in $J$,
\begin{eqnarray}
J & \propto & \int \! \! d^3x \, dx_4 \, {\Big( \frac{{\vec x}}{{x_4}^{\frac{5}{2}}} \Big)}^2
\, e^{-{\vec x}^2/2 \gamma x_4} \, {\vec x}^2 \, , \nonumber \\
& \propto &\int \! \! dx_4 \, \frac{1}{{x_4}^{\frac{3}{2}}} \, ,
\end{eqnarray}
which is singular in the ultraviolet. On the other hand, one readily checks that this term in fact is the
only one that has to be subtracted: The cubic Taylor term in the expansion of ${\partial}_s {\tilde G}^T$,
Eq.(\ref{Taylor}), already leads to a convergent contribution to the integral $J$. We now discuss the
renormalization procedure in detail, along the lines of Ref.\citep{GL89}, which we adapt to nonrelativistic
kinematics.

We first cut out a sphere ${\cal S}$ of radius $|{\cal S}| \leq \beta/2$ around the origin and decompose
the integral involving the contributions of class $C$ according to
\begin{eqnarray}
& & {\int}_{\! \! \! {\cal T}} \! \! d^{d_s+1} \! x \, {\partial}_r \Delta
 {\partial}_r \Delta \, {\partial}_s {\tilde G}^T {\partial}_s {\tilde G}^T
\nonumber \\
& & = {\int}_{\! \! \! {\cal S}} \! \! d^{d_s+1} \! x \, {\partial}_r \Delta
{\partial}_r \Delta \, {\partial}_s {\tilde G}^T {\partial}_s {\tilde G}^T
+ {\int}_{\! \! \! {{\cal T} \setminus \cal S}} \! \! d^{d_s+1} \! x \,
{\partial}_r \Delta {\partial}_r \Delta \, {\partial}_s {\tilde G}^T
{\partial}_s {\tilde G}^T \, .
\end{eqnarray}
The integral over the complement ${\cal T} \setminus {\cal S}$ of the sphere is not singular in the limit
$d_s \!\to \! 3$. In the integral over the sphere, which is divergent, we subtract the singular term
discussed above, arriving at
\begin{eqnarray}
\label{sphereDecomp}
& & \hspace*{-1cm} {\int}_{\! \! \! {\cal S}} \! \! d^{d_s+1} \! x \, {\partial}_r \Delta(x) {\partial}_r
\Delta(x) \, {\partial}_s G^T(-x) {\partial}_s G^T(-x) \nonumber \\
& = & {\int}_{\! \! \! {\cal S}} \! \! d^{d_s+1} \! x \, {\partial}_r \Delta(x) {\partial}_r \Delta(x) \,
Q_{ss}(x) \nonumber \\
& & + {\int}_{\! \! \! {\cal S}} \! \! d^{d_s+1} \! x \, {\partial}_r \Delta(x) {\partial}_r \Delta(x) \,
{\partial}_{\alpha s} G^T(-x)|_{x=0} \, {\partial}_{\beta s} G^T(-x)|_{x=0} \, x^{\alpha} \, x^{\beta} \, ,
\end{eqnarray}
where the quantity $Q_{ss}(x)$ is defined as
\begin{equation}
Q_{ss}(x) = {\partial}_s G^T(-x) {\partial}_s G^T(-x) - {\partial}_{\alpha s} G^T(-x)|_{x=0} \,
{\partial}_{\beta s} G^T(-x)|_{x=0} \,  x^{\alpha} \, x^{\beta} \, .
\end{equation}
Whereas in Eq.(\ref{sphereDecomp}) the first integral on the right hand side now is convergent, the second
integral does contain the ultraviolet singularity. The last step in the isolation of this singularity
consists in decomposing the respective integral as follows:
\begin{eqnarray}
& & \hspace*{-1cm} {\int}_{\! \! \! {\cal S}} \! \! d^{d_s+1} \! x \, {\partial}_r \Delta(x) \,
{\partial}_r \Delta(x)  \,
{\partial}_{\alpha s} G^T(-x)|_{x=0} \, {\partial}_{\beta s} G^T(-x)|_{x=0} \, x^{\alpha} \, x^{\beta}
\nonumber \\ 
& = & {\int}_{\! \! \! {\cal R}} \! \! d^{d_s+1} \! x \, {\partial}_r \Delta(x) \, {\partial}_r \Delta(x)  \,
{\partial}_{\alpha s} G^T(-x)|_{x=0} \, {\partial}_{\beta s} G^T(-x)|_{x=0} \, x^{\alpha} \, x^{\beta}
\nonumber \\
& & - {\int}_{\! \! \! {\cal R} \setminus {\cal S}} \! \! d^{d_s+1} x \, {\partial}_r \Delta(x) \,
{\partial}_r \Delta(x)  \, {\partial}_{\alpha s} G^T(-x)|_{x=0} \, {\partial}_{\beta s} G^T(-x)|_{x=0} \,
x^{\alpha} \, x^{\beta} \, .
\end{eqnarray}
The UV-singularity is contained in the integral over all Euclidean space, which can be cast into the form
\begin{eqnarray}
& & {\int}_{\! \! \! {\cal R}} \! \! d^{d_s+1} \! x \, {\partial}_r \Delta(x) \, {\partial}_r \Delta(x)  \,
{\partial}_{\alpha s} G^T(-x)|_{x=0} \, {\partial}_{\beta s} G^T(-x)|_{x=0} \, x^{\alpha} \, x^{\beta}
\nonumber \\
& = & \frac{d_s(d_s+2)}{2^{3d_s+5} \pi^{\frac{3d_s}{2}} \gamma^{\frac{3d_s+4}{2}}} \ T^{d_s+2} \,
{(\mu H)}^{\frac{d_s-2}{2}} \, { \Bigg\{ \sum_{n=1}^{\infty} \,
\frac{e^{- \mu H n \beta}}{n^{\frac{d_s+2}{2}}} \Bigg\} }^2 \, \Gamma(1-\frac{d_s}{2}) \, .
\end{eqnarray}
In the limit $d_s \!\to \! 3$ the above regularized expression is finite and takes the value
\begin{equation}
- \frac{15}{8192 {\pi}^4 \gamma^{\frac{13}{2}}} \, T^{\frac{11}{2}} \, \sqrt{\sigma} \
{ \Bigg\{ \sum_{n=1}^{\infty} \, \frac{e^{- \sigma n}}{n^{\frac{5}{2}}} \Bigg\} }^2 \, ,
\end{equation}
where we have defined the dimensionless quantity $\sigma$ as
\begin{equation}
\sigma = \mu H \beta = \frac{\mu H}{T} \, .
\end{equation}

Collecting the various contributions we arrive at the following representation for the renormalized
integral ${\bar J}$:
\begin{eqnarray}
\label{J bar}
{\bar J} & = & {\int}_{\! \! \! {\cal T}} \! \! d^4x \, \Big( {\partial}_r G^T
{\partial}_r G^T {\partial}_s {\tilde G}^T {\partial}_s {\tilde G}^T
+ 4 \, {\partial}_r \Delta \, {\partial}_r G^T {\partial}_s {\tilde G}^T
{\partial}_s {\tilde G}^T \Big)
\nonumber \\
& + & 2 {\int}_{\! \! \! {\cal T} \setminus {\cal S}} \! \! d^4x \, {\partial}_r
\Delta \, {\partial}_r \Delta \, {\partial}_s {\tilde G}^T \,
{\partial}_s {\tilde G}^T + 2 {\int}_{\! \! \! {\cal S}} \! \! d^4x \,
{\partial}_r \Delta \, {\partial}_r \Delta \, Q_{ss} \nonumber \\
& - & 2 \, {\int}_{\! \! \! {\cal R} \setminus {\cal S}} \! \! d^4 x \, {\partial}_r
\Delta \, {\partial}_r \Delta \, {\partial}_{\alpha s} G^T(-x)|_{x=0} \,
{\partial}_{\beta s} G^T(-x)|_{x=0} \,
 x^{\alpha} \, x^{\beta} \nonumber \\ 
& - & \frac{15}{4096 {\pi}^4 \gamma^{\frac{13}{2}}} \, T^{\frac{11}{2}} \, \sqrt{\sigma} \
{ \Bigg\{ \sum_{n=1}^{\infty} \, \frac{e^{- \sigma n}}{n^{\frac{5}{2}}} \Bigg\} }^2 \, .
\end{eqnarray}
Note that all terms therein are well-defined at the physical dimension $d_s=3$.

Since the various integrands only depend on the variables $r \! = \! |{\vec x}|$ and $t \! = \! x_4$, the
integrals become in fact two-dimensional,
\begin{equation}
d^4 x = 4 \pi r^2 dr \, dt \, ,
\end{equation}
and the numerical evaluation of the integral ${\bar J}$ is straightforward. A very welcome consistency check
on the numerics is provided by the fact that the result must be independent of the radius of the sphere
${\cal S}$. While more details concerning the numerical evaluation can be found in Appendix
\ref{appendix A}, in the next section we discuss the result for the function ${\bar J}={\bar J}(\sigma)$. In
particular, we consider the limit $\sigma \! \to \!0$, which is needed for the evaluation of the spontaneous
magnetization.

\section{Thermodynamics of the Ideal Ferromagnet}
\label{thermodynamics}

\begin{figure}[t]
\begin{center}
\epsfig{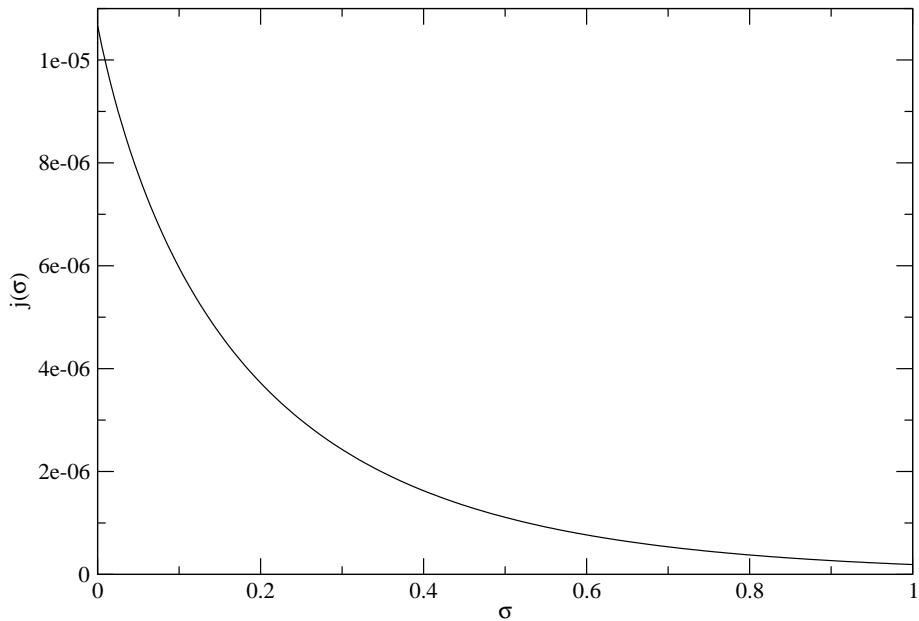}
\end{center}
{\caption{The function $j(\sigma)$, where $\sigma$ is the dimensionless parameter $\sigma = \mu H / T$.}
\label{figure3} }
\end{figure}

For dimensional reasons, the renormalized integral ${\bar J}$ can be written as
\begin{equation}
\label{expansion J}
{\bar J}(\sigma) =  T^{\frac{11}{2}} \, \frac{j(\sigma)}{\gamma^{\frac{13}{2}}} \, ,
\qquad \sigma=\frac{\mu H}{T} \, , \qquad \gamma = \frac{F^2}{\Sigma} \, ,
\end{equation}
where the quantity $j(\sigma)$ is a dimensionless function. A graph is provided in Fig.~\ref{figure3}.
In the limit $\sigma \! \to \!0$, the function can be parametrized by
\begin{equation}
\label{expansion J}
j(\sigma) = j_1 + j_2 \, \sigma + {\cal O}({\sigma}^{3/2} ) \, .
\end{equation}
The coefficients $j_1$ and $j_2$ are pure numbers given by
\begin{equation}
j_1 = 1.07 \times 10^{-5} \, , \qquad j_2 = -8  \times 10^{-5} \, .
\end{equation}
It should be noted that, in the limit $\sigma \! \to \!0$, the last two contributions in Eq.(\ref{J bar})
contain terms involving the square root $\sqrt{\sigma}$. Since they have opposite signs, however, they
cancel each other.

With the above representation for the quantity $j(\sigma)$, the final result for the low-temperature
expansion of free energy density for the ideal ferromagnet up to order $p^{11}$ takes the form
\begin{eqnarray}
\label{FreeCollectOrder11}
z = & - &\Sigma \mu H \; - \; \frac{1}{8 {\pi}^{\frac{3}{2}}
{\gamma}^{\frac{3}{2}}} \
T^{\frac{5}{2}} \sum^{\infty}_{n=1} \frac{e^{- \mu H n \beta}}{n^{\frac{5}{2}}}
\; - \; \frac{15 \,l_3}{16 {\pi}^{\frac{3}{2}} \Sigma {\gamma}^{\frac{7}{2}}} \
T^{\frac{7}{2}} \sum^{\infty}_{n=1} \frac{e^{- \mu H n \beta}}{n^{\frac{7}{2}}}
\nonumber \\
& - & \frac{105}{32 {\pi}^{\frac{3}{2}} \Sigma {\gamma}^{\frac{9}{2}}} \,
\Bigg( \frac{9 l^2_3}{2 \Sigma \gamma} - c_1 \Bigg) \ T^{\frac{9}{2}}
\sum^{\infty}_{n=1} \frac{e^{- \mu H n \beta}}{n^{\frac{9}{2}}}
\nonumber \\
& - & \frac{3 (8 l_1 + 6 l_2 + 5 l_3)}{128 {\pi}^3 {\Sigma}^2 {\gamma}^5} \ T^5
{\Bigg\{ \sum^{\infty}_{n=1} \frac{e^{- \mu H n\beta}}{n^{\frac{5}{2}}} \Bigg\}}^2
\nonumber \\
& - & \frac{945 \, d_1}{64 {\pi}^{\frac{3}{2}} \Sigma
{\gamma}^{\frac{11}{2}}} \, T^{\frac{11}{2}} \sum^{\infty}_{n=1}
\frac{e^{- \mu H n \beta}}{n^{\frac{11}{2}}}
+ \frac{10395 \, l_3 c_1}{64 {\pi}^{\frac{3}{2}} {\Sigma}^2
{\gamma}^{\frac{13}{2}}} \, T^{\frac{11}{2}} \sum^{\infty}_{n=1}
\frac{e^{- \mu H n \beta}}{n^{\frac{11}{2}}} \nonumber \\
& - & \frac{1}{2 \Sigma^2 \gamma^{\frac{9}{2}}} \, j(\mu H \beta) \, T^{\frac{11}{2}}
+ {\cal O}(T^6) \, .
\end{eqnarray}

Because the system is homogeneous, the pressure can be obtained from the temperature-dependent part of the
free energy density,
\begin{equation}
\label{Pz}
P \, = \, z_0 - z \, .
\end{equation}
Accordingly, up to order $p^{11}$, the low-temperature series for the pressure reads
\begin{equation}
\label{Pressure}
P \; = \; h_0 \, T^{\frac{5}{2}} \, + \, h_1 \, T^{\frac{7}{2}} \, + \, h_2 \,
T^{\frac{9}{2}} \, + \, h_3 \, T^5 \, + \, h_4 \, T^{\frac{11}{2}} + {\cal O}(T^6) \, ,
\end{equation}
where the coefficients $h_i$ are given by
\begin{eqnarray}
h_0 & = & \frac{1}{8 {\pi}^{\frac{3}{2}} {\gamma}^{\frac{3}{2}}} \, \sum^{\infty}_{n=1}
\frac{e^{- \mu H n \beta}}{n^{\frac{5}{2}}} \, , \nonumber \\
h_1 & = & \frac{15 \, l_3}{16 {\pi}^{\frac{3}{2}} \Sigma {\gamma}^{\frac{7}{2}}} \,
\sum^{\infty}_{n=1} \frac{e^{- \mu H n \beta}}{n^{\frac{7}{2}}} \, , \nonumber \\
h_2 & = & \frac{105}{32 {\pi}^{\frac{3}{2}} \Sigma {\gamma}^{\frac{9}{2}}} \, \Bigg(
\frac{9 l^2_3}{2 \Sigma \gamma} \, - \, c_1 \Bigg) \, \sum^{\infty}_{n=1}
\frac{e^{- \mu H n \beta}}{n^{\frac{9}{2}}} \, , \nonumber \\
h_3 & = & \frac{3 (8 l_1 + 6 l_2 + 5 l_3)}{128 {\pi}^3
{\Sigma}^2 {\gamma}^5} \, {\Bigg\{
\sum^{\infty}_{n=1} \frac{e^{- \mu H n \beta}}{n^{\frac{5}{2}}} \Bigg\}}^2 \, , \nonumber \\
h_4 & = & \frac{945}{64 {\pi}^{\frac{3}{2}} \Sigma {\gamma}^{\frac{11}{2}}} \, \Bigg(
d_1 - \frac{11 l_3 c_1}{\Sigma \gamma} \Bigg) \, \sum^{\infty}_{n=1} \frac{e^{- \mu H n \beta}}{n^{\frac{11}{2}}}
+ \frac{1}{2 \Sigma^2 \gamma^{\frac{9}{2}}} \, j \, .
\end{eqnarray}
In the limit $\sigma = \mu H / T \! \to \! 0$, these coefficients become temperature independent and the
sums reduce to Riemann zeta functions,
\begin{eqnarray}
\label{FreeCollectT(H=0)}
{\tilde h_0} & = & \frac{1}{8 {\pi}^{\frac{3}{2}} {\gamma}^{\frac{3}{2}}} \,
\zeta(\mbox{$ \frac{5}{2}$})
\, , \nonumber \\
{\tilde h_1} & = & \frac{15 \, l_3}{16 {\pi}^{\frac{3}{2}} \Sigma {\gamma}^{\frac{7}{2}}}
\, \zeta(\mbox{$ \frac{7}{2}$}) \, , \nonumber\\
{\tilde h_2} & = & \frac{105}{32 {\pi}^{\frac{3}{2}} \Sigma {\gamma}^{\frac{9}{2}}} \,
\Bigg( \frac{9 l^2_3}{2 \Sigma \gamma} \, - \, c_1 \Bigg) \,
\zeta(\mbox{$ \frac{9}{2}$}) \, , \nonumber \\
{\tilde h_3} & = & \frac{3 (8 l_1 + 6 l_2 + 5 l_3)}{128 {\pi}^3
{\Sigma}^2 {\gamma}^5} \, \zeta^2(\mbox{$ \frac{5}{2}$}) \, , \nonumber \\
{\tilde h_4} & = & \frac{945}{64 {\pi}^{\frac{3}{2}} \Sigma {\gamma}^{\frac{11}{2}}} \, \Bigg(
d_1 - \frac{11 l_3 c_1}{\Sigma \gamma} \Bigg) \, \zeta(\mbox{$ \frac{11}{2}$})
+ \frac{1}{2 \Sigma^2 \gamma^{\frac{9}{2}}} \, j_1 \, .
\end{eqnarray}
The spin-wave interaction manifests itself in the last two terms involving the coefficients ${\tilde h_3}$
and ${\tilde h_4}$. The contribution proportional to five powers of the temperature in the pressure is the
famous Dyson term. In the effective theory it originates from the two-loop graphs 10a and 10b of
Fig.~\ref{figure1}. Our main new result concerns the manifestation of the spin-wave interaction beyond
Dyson: the leading correction in the pressure is of order $T^{11/2}$. It is contained in the last term of
the coefficient ${\tilde h_4}$ and stems from the three-loop graph 11c. 

Note that all other contributions in the pressure up to order $p^{11}$ originate from one-loop graphs --
those graphs describe noninteracting magnons and merely modify the dispersion relation. In the above series
for the pressure, they involve half-integer powers of the temperature: $T^{5/2}, T^{7/2}, T^{9/2}$ and
$T^{11/2}$.

We have to point out that the sign of the Dyson term of order $T^5$ is not determined by the symmetries --
the low-energy constants $l_1, l_2$ and $l_3$ appearing in the coefficient ${\tilde h_3}$ may take positive
or negative values, depending on the specific underlying model. For the present case of the Heisenberg
model, however, Dyson has derived an explicit microscopic expression for ${\tilde h_3}$. As it turns out,
for all three types of cubic lattices, this coefficient is positive, leading to a positive contribution to
the pressure. We thus conclude that the spin-wave interaction in the ideal ferromagnet is repulsive at low
temperatures.

Remarkably, while the sign of the coefficient of order $T^5$ is not determined within the effective theory
framework, the sign of the coefficient of the subsequent interaction contribution of order $T^{11/2}$ is
unambiguously fixed: the last term in ${\tilde h_4}$ only involves the coupling constants of the
leading-order effective Lagrangian ${\cal L}^2_{eff}$ and the coefficient $j_1$ which is a pure number -- the
conditions imposed by symmetry are thus very restrictive here. Since the numerical value of $j_1$ is
positive, the corresponding contribution to the pressure is positive as well, enhancing thus the weak
repulsive interaction between spin waves at low temperatures.

Finally, let us consider the low-temperature series for the energy density $u$, for the entropy density $s$,
and for the heat capacity $c_V$ of the O(3) ferromagnet. They are readily worked out from the thermodynamic
relations
\begin{equation}
\label{Thermodynamics}
s = \frac{{\partial}P}{{\partial}T} \, , \qquad u = Ts - P \, , \qquad 
c_V = \frac{{\partial}u}{{\partial}T} = T \, \frac{{\partial}s}{{\partial}T} \, .
\end{equation}
In the limit $\sigma \! \to \! 0$, we obtain
\begin{eqnarray}
u & = & \mbox{$ \frac{3}{2}$} \, {\tilde h_0} \, T^{\frac{5}{2}} \,
+ \, \mbox{$ \frac{5}{2}$} \, {\tilde h_1} \, T^{\frac{7}{2}} \,
+ \, \mbox{$ \frac{7}{2}$} \, {\tilde h_2} \, T^{\frac{9}{2}} \,
+ \, 4 {\tilde h_3} \, T^5
+ \mbox{$ \frac{9}{2}$} \,  {\tilde h_4} \, T^{\frac{11}{2}}
+ {\cal O}(T^6) \, ,
\nonumber \\
s & = & \mbox{$ \frac{5}{2}$} \, {\tilde h_0} \, T^{\frac{3}{2}} \,
+ \, \mbox{$ \frac{7}{2}$} \, {\tilde h_1} \, T^{\frac{5}{2}} \,
+ \, \mbox{$ \frac{9}{2}$} \, {\tilde h_2} \, T^{\frac{7}{2}} \,
+ \, 5 {\tilde h_3} \, T^4
+ \,  \mbox{$ \frac{11}{2}$} \, {\tilde h_4} \, T^{\frac{9}{2}}
+ {\cal O}(T^5) \, ,
\nonumber \\
c_V & = & \mbox{$ \frac{15}{4}$} \, {\tilde h_0} \, T^{\frac{3}{2}}
\, + \, \mbox{$ \frac{35}{4}$} \, {\tilde h_1} \, T^{\frac{5}{2}}
\, + \, \mbox{$ \frac{63}{4}$} \, {\tilde h_2} \, T^{\frac{7}{2}} \,
+ \, 20 {\tilde h_3} \, T^4
+ \, \mbox{$ \frac{99}{4}$} \, {\tilde h_4} \, T^{\frac{9}{2}}
+ {\cal O}(T^5) \, .
\end{eqnarray}
Again, the correction to Dyson's result is contained in the respective last terms in the above series
involving the coefficient ${\tilde h_4}$.

\section{Spontaneous Magnetization: Effective Framework versus Condensed Matter Literature}
\label{spontaneousMagnetization}

We now turn to the discussion of the general structure of the low-temperature series for the spontaneous
magnetization of the ideal ferromagnet. While this problem has attracted more than a hundred authors over the
last few decades, to the best of our knowledge, a rigorous and fully systematic calculation of higher-order
corrections to the Dyson term has never been achieved. Before we review the relevant results in the
literature, let us analyze the problem within the systematic effective field theory framework.

With the expression for the free energy density (\ref{FreeCollectOrder11}), the low-temperature expansion
for the spontaneous magnetization
\begin{equation}
\Sigma(T)\, = \, - \lim_{H \to 0} \frac{\partial z}{\partial(\mu H)}
\end{equation}
of the O(3) ferromagnet, up to order $T^{9/2}$, takes the form
\begin{equation}
\frac{\Sigma(T)}{\Sigma} = 1 - {\alpha}_0 T^{\frac{3}{2}} - {\alpha}_1 T^{\frac{5}{2}}
- {\alpha}_2 T^{\frac{7}{2}} - {\alpha}_3 T^4 - {\alpha}_4 T^{\frac{9}{2}} + {\cal O}(T^5) \, .
\end{equation}
The coefficients $\alpha_i$ are independent of the temperature and given by
\begin{eqnarray}
\label{SigmaCollectT(H=0)}
{\alpha}_0 & = & \frac{1}{8 {\pi}^{\frac{3}{2}} \Sigma {\gamma}^{\frac{3}{2}}} \,
\zeta(\mbox{$ \frac{3}{2}$}) \, , \nonumber \\
{\alpha}_1 & = & \frac{15 \, l_3}{16 {\pi}^{\frac{3}{2}} {\Sigma}^2
{\gamma}^{\frac{7}{2}}} \, \zeta(\mbox{$ \frac{5}{2}$}) \, , \nonumber \\
{\alpha}_2 & = & \frac{105}{32 {\pi}^{\frac{3}{2}} {\Sigma}^2 {\gamma}^{\frac{9}{2}}} \,
\Bigg( \frac{9 l^2_3}{2 \Sigma \gamma} \, - \, c_1 \Bigg) \,
\zeta(\mbox{$ \frac{7}{2}$}) \, , \nonumber \\
{\alpha}_3 & = & \frac{3 (8 l_1 + 6 l_2 + 5 l_3)}{64 {\pi}^3 {\Sigma}^3 {\gamma}^5}\
\zeta(\mbox{$ \frac{5}{2}$}) \, \zeta(\mbox{$ \frac{3}{2}$}) \, , \nonumber \\
{\alpha}_4 & = & \frac{945}{64 {\pi}^{\frac{3}{2}} {\Sigma}^2 {\gamma}^{\frac{11}{2}}} \,
\Bigg( d_1 - \frac{11 l_3 c_1}{\Sigma \gamma} \Bigg) \, \zeta(\mbox{$ \frac{9}{2}$})
- \frac{1}{2 \Sigma^3 \gamma^{\frac{9}{2}}} \, j_2 \, .
\end{eqnarray}
Up to order $T^4$, we reproduce Dyson's series. In the effective Lagrangian framework, the famous
interaction term of order $T^4$ in the spontaneous magnetization originates from the two-loop graphs 10a and
10b which involve vertices from the next-to-leading order Lagrangian ${\cal L}^4_{eff}$. Note that there is
no interaction term of order $T^3$ in the above series -- for many years such a spurious term has haunted
the condensed matter literature.\footnote{The first encounter with this spurious term seems to date back to
the year 1958 (see Ref.~\citep{Man58}). It then continued haunting the literature over a period of at least
25 years until 1983 when it was last sighted in Ref.~\citep{KG83}.}

Our main new result concerns the leading correction to Dyson's term, which originates from the three-loop
graph 11c. The correction in the spontaneous magnetization is of order $T^{9/2}$. Remarkably, the
corresponding coefficient -- the last term in ${\alpha}_4$ -- does not involve any higher-order low-energy
constants. It only involves $\Sigma$ and $F$, as well as the coefficient $j_2$, which is a pure number
determined by the symmetries of the underlying Heisenberg model. Since the coefficient $j_2$ is negative,
this contribution has the same sign as the Dyson coefficient ${\alpha}_3$. The effect of the three-loop
contribution is thus to enhance the weak spin-wave interaction found by Dyson.

Apart from these two interaction terms of order $T^4$ and $T^{9/2}$, respectively, all other
temperature-dependent contributions to the spontaneous magnetization originate from one-loop graphs, which
describe noninteracting magnons. They merely modify the dispersion relation or -- as Dyson expressed it
\citep{Dys56} -- they merely arise {\it from the discreteness of the lattice, are easy to calculate and are
not of any theoretical interest}. In the above series for the spontaneous magnetization, they involve
half-integer powers of the temperature: $T^{3/2}, T^{5/2},T^{7/2}$ and $T^{9/2}$.

It should be pointed out that the contribution of order $T^{9/2}$ contains two parts: The first term
in the coefficient ${\alpha}_4$ is due to two one-loop graphs associated with noninteracting magnons. The
second term is due to a three-loop graph and represents the dominant spin-wave interaction term beyond
Dyson. Note that the Dyson coefficient ${\alpha}_3$, on the other hand, exclusively involves an interaction
part.

Here comes the appropriate place to compare our results with the condensed matter literature. Indeed,
several authors -- most notably, Dyson himself -- also have discussed the structure of the low-temperature
series for the spontaneous magnetization beyond order $T^4$. We make our comparison along four lines of
observations.

Our first observation is that all published calculations or estimates of higher-order interaction terms
\citep{Dys56,MT65b,Cha01,Ach11} apparently failed to identify the dominant $T^{9/2}$-correction to the Dyson
term in the spontaneous magnetization.

The second observation is that there appears to be consensus in the literature on how graphs related to the
{\it two}-spin-wave problem should manifest themselves beyond $T^4$. Dyson classified his terms according to
the quantity $F$, where $F$ is the {\it number of independent particles which are concerned in the
interactions which the particular term describes} \citep{Dys56}. For $F=2$, which is referred to as
the {\it two-spin-wave problem} in Ref.~\citep{MT65b}, the corresponding corrections are expected to show
up at order $T^5$ according to Refs.~\citep{MT65b,Cha01,Ach11} -- hence these authors seem to agree on that
the dominant correction to the Dyson term should be of order $T^5$ in the spontaneous magnetization.
However, this claim is not correct -- it is in contradiction with the fully systematic effective field
theory analysis which has demonstrated that the dominant correction sets in at order $T^{9/2}$.

The third observation concerns the {\it three}-spin-wave problem, i.e. the effect of interaction terms with
$F=3$. Dyson identified two such contributions -- formulae (128) and (130) in his second article of
Ref.~\citep{Dys56} -- and showed that in the spontaneous magnetization these are of order $T^{13/2}$ and
$T^5$, respectively. In the article by Morita and Tanaka \citep{MT65b}, however, it is claimed that the
{\it three}-spin-wave problem starts manifesting itself at order $T^{13/2}$, missing thus the term of order
$T^5$. 

Finally, the fourth observation is that the only place in the literature where a Feynman diagram displaying
the cateye structure of graph 11c seems to have appeared, is in the more recent article by Chang
\citep{Cha01}. However, he concludes that interactions originating from such a diagram start showing up only
at order $T^{15/2}$ in the spontaneous magnetization. This claim, again, is erroneous, as it contradicts the
fully systematic effective theory analysis which has demonstrated that the leading term originating from a
cateye graph is of order $T^{9/2}$.

In view of the quite impressive collection of temperature powers established over the years, one may easily
get confused -- after all, may some of these temperature powers, again, merely be spurious? One would
certainly like to gain some deeper insight into the general structure of the low-temperature series for the
spontaneous magnetization beyond the leading correction to the Dyson term. Let us therefore address the
problem in a fully systematic way within the effective field theory framework -- first on the level of the
free energy density. 

\begin{figure}[t]
\begin{center}
\epsfig{file=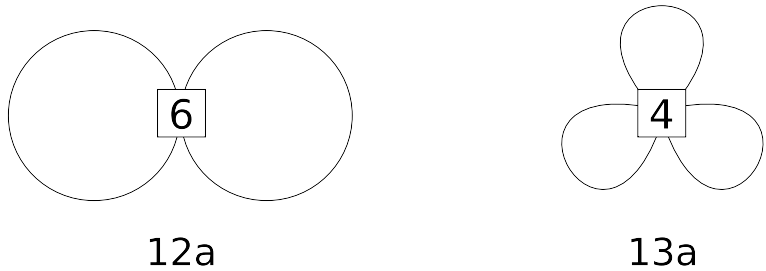,width=10cm}
\end{center}
{\caption{Two Feynman graphs related to the low-temperature expansion of the partition function for a
ferromagnet at order $p^{12}$ and $p^{13}$ in dimension $d$=3+1. The numbers attached to the vertices refer
to the piece of the effective Lagrangian they come from. Note that there are further Feynman graphs of order
$p^{12}$ and $p^{13}$ which we have not displayed.}
\label{figure4} }
\end{figure} 

Indeed, it is quite easy to see that corrections due to the spin-wave interaction continue to proceed in
steps of $T^{1/2}$. In Fig.~\ref{figure4} we have displayed some of the relevant higher-order graphs which
contribute beyond order $p^{11}$ or, equivalently, beyond $T^{11/2}$. At order $T^6$ in the free energy
density, the two-loop graph 12a with an insertion from ${\cal L}^6_{eff}$ contributes, while at order
$T^{13/2}$ the three-loop graph 13a with a vertex from ${\cal L}^4_{eff}$ is relevant.

We may classify the graphs of the effective theory according to the number of loops they contain and discuss
the various contributions at the level of the spontaneous magnetization. Interactions related to two-loop
diagrams start manifesting themselves through the Dyson term of order $T^4$ and then proceed in integer
steps of $T$ -- these are the two-loop graphs with successive insertions from
${\cal L}^4_{eff}, {\cal L}^6_{eff}, {\cal L}^8_{eff}, \dots$, giving rise to terms of order
$T^4, T^5, T^6, \dots$ in the spontaneous magnetization. Interactions related to three-loop diagrams start
showing up at order $T^{9/2}$. They give rise to the dominant correction to Dyson's result, and then also
proceed in steps of $T$ -- in the effective theory these correspond to the three-loop graphs with successive
insertions from ${\cal L}^2_{eff}, {\cal L}^4_{eff},  {\cal L}^6_{eff}, \dots$, leading to half-integer powers
of the temperature: $T^{9/2}, T^{11/2}, T^{13/2}, \dots$ Four-loop interactions are expected to enter the game
at order $T^6$. They will continue contributing to the spontaneous magnetization in ascending powers of $T$
through terms of order $T^6, T^7, T^8, \dots$

Accordingly, the low-temperature expansion for the spontaneous magnetization of the ideal ferromagnet
exhibits the following general structure:
\begin{equation}
\frac{\Sigma(T)}{\Sigma} =  1 - {\alpha}_0 T^{\frac{3}{2}} - {\alpha}_1 T^{\frac{5}{2}}
- {\alpha}_2 T^{\frac{7}{2}} - {\bf {\alpha}_3 T^4} - {\bf {\alpha}_4 T^{\frac{9}{2}}} - {\bf {\alpha}_5 T^5}
- {\bf {\alpha}_6 T^{\frac{11}{2}}}+ {\cal O}({\bf T^6}) \, .
\end{equation}
Note that we have highlighted all terms which are related to the spin-wave interaction. We thus realize that
the various published temperature powers as such are not in contradiction with the effective field theory
prediction -- sooner or later, the effective expansion will hit them all. The point is, however, that there
are two gaps in the multitude of published temperature powers -- interaction terms of order $T^{9/2}$ and
$T^{11/2}$, to the best of our knowledge, have never been identified.

One may say that the effect of the spin-wave interaction on the low-temperature series for the spontaneous
magnetization of an ideal ferromagnet is quite peculiar. On the one hand, the interaction starts manifesting
itself only at order $T^4$, i.e. far beyond the Bloch term of order $T^{3/2}$. On the other hand,
subsequent interaction corrections closely follow the term of order $T^4$, starting with $T^{9/2}$ and then
proceeding in steps of $T^{1/2}$.

\section{Conclusions}
\label{conclusions}

The question of how the spin-wave interaction manifests itself in the low-temperature expansion of the
spontaneous magnetization of an ideal ferromagnet, has a long history. Early attempts that ended up with
temperature powers of order $T^{7/4}$ and $T^2$ turned out to be wrong altogether, as shown by Dyson's
rigorous analysis which demonstrated that the spin-wave interaction sets in only at order $T^4$.

After Dyson's rather complicated analysis, there emerged an active phase of research in which many authors
tried to derive the $T^4$-term in the spontaneous magnetization in a more accessible manner. It is
interesting to note that this problem -- using conventional condensed matter methods -- indeed seems to be
nontrivial, as various authors all of a sudden ended up with an interaction term of order $T^3$ in the
spontaneous magnetization, thereby contradicting Dyson. As it turned out, these attempts were wrong
altogether, as they were plagued with errors originating from the approximate methods used.

We would like to stress that the fully systematic effective Lagrangian method, based on symmetry
considerations, does not display any such defects. Rather, the existence of Dyson's $T^4$-term -- and, at
the same time, the absence of a $T^3$-term -- is an immediate consequence of the underlying symmetries
inherent in the ideal ferromagnet.

In his collection of selected papers that appeared in 1996, Dyson comments \citep{Dys96}: {\it "After 1966,
the subject of spin-wave interactions went into a long sleep."} \footnote{In 1966 Keffer's comprehensive
review on spin waves appeared (Ref.~\citep{Kef66}).} Still, so it seems, sporadically over time, the
peaceful sleep has been interrupted, as several authors were attracted by the problem of how the general
structure of the series for the spontaneous magnetization of an ideal ferromagnet beyond the Dyson term
should look like. Remarkably, not all of the various findings are consistent with one another. It was our
motivation to solve this paradox, making use of the systematic and model-independent method of effective
Lagrangians.

As we have demonstrated in the present study, it is rather straightforward to go beyond Dyson's analysis by
considering three-loop effects in the effective field theory. Still, the explicit evaluation at order
$p^{11}$ is quite non-trivial as it involves the renormalization and subsequent numerical evaluation of a
three-loop graph, which is proportional to an integral over a product of four thermal propagators, each one
of them involving an infinite sum. The corresponding interaction term beyond Dyson is completely fixed by
the symmetries of the leading-order effective Lagrangian ${\cal L}^2_{eff}$ -- lattice anisotropies, showing
up at higher orders in the effective Lagrangian, do not affect this result. What is quite remarkable is the
fact that all previous attempts to go beyond Dyson apparently have failed to correctly identify this
interaction term of order $T^{9/2}$ in the spontaneous magnetization of an ideal ferromagnet.

We have also discussed the origin and the structure of even higher-order corrections in the low-temperature
expansion of the spontaneous magnetization, pointing out that they continue to proceed in steps of $T^{1/2}$
beyond the contribution of order $T^{9/2}$. Again, earlier attempts to gain insight into the general
structure of this series were incorrect.

The present study has thus solved -- once and for all -- the problem of how the spin-wave interaction in an
ideal ferromagnet manifests itself in low-temperature expansion of the spontaneous magnetization beyond the
Dyson term.

Hopefully, we have convinced the reader that the effective Lagrangian technique does not merely consist in
rederiving known results or in rephrasing condensed matter problems in another language -- rather, in many
cases as in the present one, it clearly proves to be more powerful than conventional condensed matter
methods, allowing one to go to higher orders of the low-temperature expansion in a controlled and systematic
manner. In view of the many articles that have dealt with the problem of the manifestation of the spin-wave
interaction in an ideal ferromagnet at low temperatures -- concerning both the extension of Dyson's series to
higher orders and the confusion regarding the spurious $T^3$-term -- it is quite striking how efficiently
the effective theory analysis settles all these questions in a conclusive way.

We do not claim to have contributed to the actual experimental situation -- spin-wave interactions in a
ferromagnet are very weak. Above all, there are many interactions in addition to the exchange interaction in
a {\it real} ferromagnet, which would also have to be accounted for in a more realistic approach. While this
would perfectly be feasible within the effective Lagrangian framework, here we have restricted ourselves to
the simple model of the {\it ideal} ferromagnet -- after all, it is for this system where corrections to
Dyson's result have been derived over the years.

Although there is no such object as a perfectly ideal ferromagnet in nature, still, the "clean" ideal
ferromagnet could be investigated in a numerical simulation of the Heisenberg model and the existence of the
$T^{9/2}$-term in the low-temperature expansion of the spontaneous magnetization might be verified this way.

\section*{Acknowledgments}

It is really a pleasure to thank Professor Freeman J. Dyson for correspondence and for carefully reading
the manuscript. Thanks also to H. Leutwyler and U.-J. Wiese for stimulating discussions. This work is
supported by CONACYT grant No. 50744-F.

\begin{appendix}

\section{Numerical Evaluation of the Cateye Graph}
\label{appendix A}

To numerically evaluate the integral ${\bar J}$ defined in Eq.~(\ref{J bar}), we introduce the dimensionless
variables $\eta$ and $\xi$,
\begin{equation}
\eta = T x_4 \, , \qquad \xi = \frac{1}{2} \sqrt{\frac{T}{\gamma}} \,
|{\vec x}| \, .
\end{equation}
In the integrals over the torus which involve quartic and triple sums -- the first two terms in
Eq.~(\ref{J bar}) -- we first integrate over all three-dimensional space, ending up with one-dimensional
integrals in the variable $\eta$. For the quartic sum we obtain
\begin{eqnarray}
\label{quartic}
& & {\int}_{\! \! \! {\cal T}} \! \! d^4 x \, {\partial}_r G^T(x) \,
{\partial}_r G^T(x) \, {\partial}_s G^T(-x) \, {\partial}_s 
G^T(-x) \nonumber \\
& & = \frac{15}{2048 {\pi}^{9/2} {\gamma}^{13/2}} \, T^{\frac{11}{2}} \,
{\int}^{1/2}_{\! \! \! -1/2} \! d \eta \, \sum^{\infty}_ {n_1 \dots n_4 =1} \,
e^{-{\sigma}(n_1 + n_2 + n_3 + n_4)} \ Q(\eta,n_1,n_2,n_3,n_4) \, , \nonumber \\
& & Q(\eta,n_1,n_2,n_3,n_4) = \frac{{\Bigg( \frac{1}{\eta + n_1}
+ \frac{1}{\eta + n_2} +\frac{1}{-\eta + n_3} +\frac{1}{-\eta + n_4}
\Bigg)}^{-7/2}}
{{\Big( (\eta + n_1)(\eta + n_2)(-\eta + n_3)(-\eta + n_4)\Big)}^{5/2}} \, ,
\end{eqnarray}
while for the triple sum we get
\begin{eqnarray}
\label{triple}
& & {\int}_{\! \! \! {\cal T}} \! \! d^4 x \, {\partial}_r \Delta(x) \,
{\partial}_r G^T(x) \, {\partial}_s G^T(-x) \,
{\partial}_s G^T(-x) \nonumber \\
& & = \frac{15}{2048 {\pi}^{9/2} {\gamma}^{13/2}} \, T^{\frac{11}{2}} \,
{\int}^{1/2}_{\! \! \! 0} \! d \eta \, \sum^{\infty}_ {n_2 \dots n_4 =1} \,
e^{-{\sigma}(n_2 + n_3 + n_4)} \ Q(\eta,0,n_2,n_3,n_4) \, , \nonumber \\
& & Q(\eta,0,n_2,n_3,n_4) = \frac{{\Bigg( \frac{1}{\eta} +\frac{1}{\eta + n_2}
+ \frac{1}{-\eta + n_3} +\frac{1}{-\eta+n_4} \Bigg)}^{-7/2}}
{{\Big( \eta(\eta + n_2)(-\eta + n_3)(-\eta+n_4)\Big)}^{5/2}} \, ,
\end{eqnarray}
with
\begin{equation}
\sigma \equiv \frac{\mu H}{T} \, , \qquad \gamma \equiv \frac{F^2}{\Sigma} \, . 
\end{equation}
Note that for the triple sums the integration over $\eta$ only extends over the interval $[0, \frac{1}{2}]$,
due to the $\Theta$-function contained in the zero-temperature propagator $\Delta(x)$.

The quantities $Q(\eta,n_1,n_2,n_3,n_4)$ and $Q(\eta,0,n_2,n_3,n_4)$ depend in a nontrivial manner on the
summation variables. The slowest convergence for the entire expressions Eq.~(\ref{quartic}) and
Eq.~(\ref{triple}) is observed for the case $\sigma=0$, where no exponential damping occurs. We have
performed the numerical summation in a "Cartesian" way. We first define the vector
${\vec N_i} = (n_1, n_2, n_2, n_4)$.  The first partial sum $S_1$ in the quartic series simply corresponds
to the combination ${\vec N_1} = (1,1,1,1)$ of indices. The second partial sum $S_2$ then contains all
combinations of indices in the vector ${\vec N_2}$ with at least one index equal to two:
$(2,1,1,1), \dots, (2,2,2,2)$, etc. For large values of $i$ and for $\sigma=0$, the partial sums $S_i$
converge according to $1/{S_i}^{5/2}$. Proceeding in an analogous manner for the triple sums, one obtains
the same asymptotic behavior.

Expressions suitable for the numerical evaluation of the remaining three integrals of Eq.(\ref{J bar})
involving double sums are
\begin{eqnarray}
& & {\int}_{\! \! \! {{\cal T} \setminus {\cal S}}} \! \! d^4 x \, {\partial}_r
\Delta(x) \, {\partial}_r \Delta(x) \, {\partial}_s G^T(-x) \,
{\partial}_s G^T(-x) \nonumber \\
& & = \frac{1}{128 {\pi}^5 {\gamma}^{13/2}} \, T^{\frac{11}{2}} \,
{\int}_{\! \! \! \! 0}^{S} \! d \eta \, {\int}_{\! \! \! \! \sqrt{S^2-{\eta}^2}}^{\infty} \! d \xi \, {\xi}^6 \,
\sum^{\infty}_ {n_1, n_2 =1} \, e^{-{\sigma}(n_1 + n_2)} \ P(\xi,\eta,n_1,n_2) \, ,
 \nonumber \\
& & P(\xi,\eta,n_1,n_2) = \frac{e^{ -{\xi}^2 \Big( \frac{2}{\eta} + \frac{1}{-\eta + n_1}
+ \frac{1}{-\eta + n_2} \Big)}}
{{\Big\{ {\eta}^2(-\eta + n_1)(-\eta + n_2)\Big\}}^{5/2}} \, ,
\end{eqnarray}
\begin{eqnarray}
& & {\int}_{\! \! \! {\cal S}} \! \! d^4x \, {\partial}_r \Delta(x) \, {\partial}_r \Delta(x) \,
Q_{ss}(x) \\
& & = \frac{1}{128 {\pi}^5 {\gamma}^{13/2}} \, T^{\frac{11}{2}} \,
{\int}_{\! \! \! \! 0}^S \! d \eta \, {\int}_{\! \! \! \! 0}^{\sqrt{S^2-{\eta}^2}} \! d \xi \, {\xi}^6 \,
\sum^{\infty}_ {n_1, n_2 =1} \, e^{-{\sigma}(n_1 + n_2 + 2\eta)} \ Q(\xi,\eta,n_1,n_2,\sigma) \, , \nonumber
\end{eqnarray}
with 
\begin{equation}
Q(\xi,\eta,n_1,n_2,\sigma) = \frac{e^{ -{\xi}^2 \Big( \frac{2}{\eta} + \frac{1}{-\eta + n_1}
+ \frac{1}{-\eta + n_2} \Big)} \Big[ 
\frac{e^{2 \eta \sigma}}{ { \{ ( -\eta + n_1)(-\eta + n_2) \} }^{5/2} }
- \frac{e^{{\xi}^2 ( \frac{1}{-\eta + n_1} + \frac{1}{-\eta + n_2} ) }}{n_1^{5/2} n_2^{5/2}}
 \Big] }
{{\eta}^5}\, ,
\end{equation}
and finally,
\begin{eqnarray}
& & {\int}_{\! \! \! {\cal R} \setminus {\cal S}} \! \! d^4 x \, {\partial}_r \Delta(x) \,
{\partial}_r \Delta(x) \, {\partial}_{s \alpha} G^T(-x)|_{x=0} \,
x^{\alpha} \, {\partial}_{s \beta} G^T(-x)|_{x=0} \, x^{\beta}
\nonumber \\
& & = \ \frac{1}{128 {\pi}^5 {\gamma}^{13/2}} \, T^{\frac{11}{2}} \,
{\int}_{\! \! \! \! S}^{\infty} \! d \eta {\int}_{\! \! \! \! 0}^{\infty} \! d \xi \, {\xi}^6 \,
\sum^{\infty}_ {n_1, n_2 =1} \, e^{-\sigma(n_1 + n_2 + 2\eta)} \ R(\xi,\eta,n_1,n_2)
\nonumber \\
& & + \ \frac{1}{128 {\pi}^5 {\gamma}^{13/2}} \, T^{\frac{11}{2}} \,
{\int}_{\! \! \! \! 0}^S \! d \eta {\int}_{\! \! \! \! \sqrt{S^2-{\eta}^2}}^{\infty} \! d \xi \, {\xi}^6 \,
\sum^{\infty}_ {n_1, n_2 =1} \, e^{-\sigma(n_1 + n_2 + 2\eta)} \ R(\xi,\eta,n_1,n_2) \, ,
\nonumber \\
& & R(\xi,\eta,n_1,n_2) = \frac{e^{ -2{\xi}^2/\eta}}
{{\Big\{ {\eta}^2 n_1 n_2 \Big\}}^{5/2}} \, .
\end{eqnarray}
It is understood that in the above integrals the radius of the sphere is chosen as $S=\frac{1}{2}$. For
large values of $i$ and for $\sigma=0$, the partial sums $S_i$ related to the above three expressions
involving double sums also converge according to $1/{S_i}^{5/2}$.

\end{appendix}


\begin{thebibliography}{1}

\bibitem[Bloch (1930)]{Blo30}
F.\ Bloch, Z.\ Phys. \textbf{61}, 206 (1930).

\bibitem[Kramers (1936)]{Kra36}
H.\ A.\ Kramers, Commun.\ Kamerlingh Onnes Lab.\ Univ.\ Leiden, Suppl.\ {\bf 22}, 83 (1936).

\bibitem[Opechowski (1937)]{Ope37}
W.\ Opechowski, Physica (Amsterdam) \textbf{4}, 715 (1937).

\bibitem[Schafroth (1954)]{Sch54}
M.\ R.\ Schafroth, Proc.\ R.\ Soc.\ London, Ser.\ A \textbf{67}, 33 (1954).

\bibitem[van Kranendonk (1955)]{Kra55}
J.\ van Kranendonk, Physica (Amsterdam) \textbf{21}, 81 (1955); \textbf{21}, 749 (1955); \textbf{21}, 925
(1955).

\bibitem[Dyson (1956)]{Dys56}
F.\ J.\ Dyson, Phys.\ Rev.\ \textbf{102}, 1217 (1956); \textbf{102}, 1230 (1956).

\bibitem[Morita (1958)]{Mor58}
T.\ Morita, Prog.\ Theor.\ Phys. \textbf{20}, 614, 728 (1958).

\bibitem[Oguchi (1960)]{Ogu60}
T.\ Oguchi, Phys.\ Rev.\ \textbf{117}, 117 (1960).

\bibitem[Keffer and Loudon (1961)]{KL61}
F.\ Keffer and R.\ Loudon, J.\ Appl.\ Phys.\ (Suppl.) \textbf{32}, 2S (1961).

\bibitem[Szaniecki (1961)]{Sza61}
J.\ Szaniecki, Acta Phys.\ Polon.\ \textbf{20}, 983 (1961).

\bibitem[Zittartz (1965)]{Zit65}
J.\ Zittartz, Z. Phys. \textbf{184}, 506 (1965).

\bibitem[Wallace (1967)]{Wal67}
D.\ C.\ Wallace, Phys.\ Rev.\ \textbf{153}, 547 (1967).

\bibitem[Vaks et al. (1968)]{VLP68}
V.\ G.\ Vaks, A.\ I.\ Larkin and S.\ A.\ Pikin, Sov.\ Phys.\ JETP \textbf{26}, 188 (1968).

\bibitem[Cooke and Hahn (1970)]{CH70}
J.\ F.\ Cooke and H.\ H.\ Hahn, Phys.\ Rev.\ B \textbf{1}, 1243 (1970).

\bibitem[Szaniecki (1974)]{Sza74}
J.\ Szaniecki, J.\ Phys. C.: Solid State Phys.\ \textbf{7}, 4113 (1974).

\bibitem[Yang and Wang (1975)]{YW75}
D.\ H.\ Yang  and Y.\ Wang, Phys.\ Rev.\ B \textbf{12}, 1057 (1975).

\bibitem[Rastelli and Lindgard (1979)]{RL79}
E.\ Rastelli  and P.\ A.\ Lindgard,  J.\ Phys. C.: Solid State Phys. \textbf{12}, 1899 (1979).

\bibitem[Dyson (1996)]{Dys96}
F.\ Dyson, {\it Selected papers of Freeman Dyson with commentary}, American Mathematical Society (1996).

\bibitem[Mannari (1958)]{Man58}
I.\ Mannari, Prog.\ Theor.\ Phys.\ \textbf{19}, 201 (1958).

\bibitem[Brout and Haken (1960)]{BH60}
R.\ Brout and H.\ Haken, Bull.\ Am.\ Phys.\ Soc.\ \textbf{5}, 148 (1960).

\bibitem[Englert (1960)]{Eng60}
F.\ Englert, Phys.\ Rev.\ Lett.\ \textbf{5}, 102 (1960).

\bibitem[Tahir-Kheli and ter Haar (1962a)]{TH62a}
R.\ A.\ Tahir-Kheli and D.\ ter Haar, Phys.\ Rev.\ \textbf{127}, 88, (1962).

\bibitem[Stinchcombe et al. (1963)]{SHEB63}
R.\ B.\ Stinchcombe, G.\ Horwitz, F.\ Englert and R.\ Brout, Phys.\ Rev.\ \textbf{130}, 155 (1963).

\bibitem[Callen (1963)]{Cal63}
H.\ B.\ Callen, Phys.\ Rev.\ \textbf{130}, 890 (1963).

\bibitem[Oguchi and Honma (1963)]{OH63}
T.\ Oguchi and A.\ Honma, J.\ Appl.\ Phys.\ \textbf{34}, 1153 (1963).

\bibitem[Kuehnel (1969)]{Kue69}
A.\ K\" uhnel, J.\ Phys. C (Solid St.\ Phys.) \textbf{2}, 711 (1969).

\bibitem[Tahir-Kheli (1970)]{Tah70}
R.\ A.\ Tahir-Kheli, Phys.\ Rev.\ B \textbf{1}, 3163 (1970).

\bibitem[Coutinho Filho and Fittipaldi (1973)]{CF73}
M.\ D.\ Coutinho Filho and I.\ P.\ Fittipaldi, Phys.\ Rev.\ B \textbf{7}, 4941 (1973).

\bibitem[Kumar and Gupta (1983)]{KG83}
A.\ Kumar and A.\ K.\ Gupta, Phys.\ Rev.\ B \textbf{28}, 3968 (1983).

\bibitem[Tahir-Kheli and ter Haar (1962b)]{TH62b}
R.\ A.\ Tahir-Kheli and D.\ ter Haar, Phys.\ Rev.\ \textbf{127}, 95 (1962).

\bibitem[Ortenburger (1964)]{Ort64}
I.\ Ortenburger, Phys.\ Rev.\ \textbf{136}, A 1374 (1964).

\bibitem[Morita and Tanaka (1965)]{MT65a}
T.\ Morita and T.\ Tanaka, Phys.\ Rev.\ \textbf{137}, A 648 (1965); \textbf{138}, A1403 (1965).

\bibitem[Keffer (1966)]{Kef66}
F. Keffer, {\it Spin Waves}, in {\it Encyclopedia of Physics -- Ferromagnetism}, edited by S. Fl\"ugge and
H. P. J. Wijn (Springer, Berlin, 1966), Vol. 18-2, p. 1.

\bibitem[Loly (1987)]{Lol87}
P.\ D.\ Loly, J.\ Can.\ Phys.\ \textbf{65}, 1272 (1987).

\bibitem[Czachor and Holas (1990)]{CH90}
A.\ Czachor and A.\ Holas, Phys.\ Rev.\ B \textbf{41}, 4674 (1990).

\bibitem[Hofmann (2002)]{Hof02}
C.\ P.\ Hofmann, Phys.\ Rev.\ B \textbf{65}, 094430 (2002).

\bibitem[Weinberg (1979)]{Wei79}
S.\ Weinberg, Physica A \textbf{96}, 327 (1979).

\bibitem[Gasser and Leutwyler (1985)]{GL85}
J.\ Gasser and H.\ Leutwyler, Ann.\ Phys.\ (N.Y.) \textbf{158}, 142 (1984); Nucl.\ Phys.\ B \textbf{250},
465 (1985).

\bibitem[Leutwyler (1994a)]{Leu94a}
H.\ Leutwyler, Phys.\ Rev.\ D \textbf{49}, 3033 (1994).

\bibitem[Roman and Soto (1999a)]{RS99a}
J.\ M.\ Rom\'an and J.\ Soto, Int.\ J.\ Mod.\ Phys.\ B \textbf{13}, 755 (1999).

\bibitem[Hofmann (1999a)]{Hof99a}
C.\ P.\ Hofmann, Phys.\ Rev.\ B \textbf{60}, 388 (1999).

\bibitem[Roman and Soto (1999b)]{RS99b}
J.\ M.\ Rom\'an and J.\ Soto, Ann.\ Phys.\ \textbf{273}, 37 (1999).

\bibitem[Hasenfratz and Leutwyler (1990)]{HL90}
P.\ Hasenfratz and H.\ Leutwyler, Nucl.\ Phys.\ B \textbf{343}, 241 (1990).

\bibitem[Hasenfratz and Niedermayer (1991)]{HN91}
P.\ Hasenfratz and F.\ Niedermayer, Phys.\ Lett.\ B \textbf{268}, 231 (1991).

\bibitem[Hasenfratz and Niedermayer (1993)]{HN93}
P.\ Hasenfratz and F.\ Niedermayer, Z. Phys.\ B \textbf{92}, 91 (1993).

\bibitem[Bietenholz (1993)]{Bie93}
W.\ Bietenholz, Helv.\ Phys.\ Acta \textbf{66}, 633 (1993).

\bibitem[Hofmann (1999b)]{Hof99b}
C.\ P.\ Hofmann, Phys.\ Rev.\ B \textbf{60}, 406 (1999).

\bibitem[Hofmann (2010)]{Hof10}
C.\ P.\ Hofmann, Phys.\ Rev.\ B \textbf{81}, 014416 (2010).

\bibitem[Roman and Soto (2000)]{RS00}
J.\ M.\ Rom\'an and J.\ Soto, Phys.\ Rev.\ B \textbf{62}, 3300 (2000).

\bibitem[Kaempfer et al. (2005)]{KMW05}
F.\ K\"ampfer, M.\ Moser and U.-J.\ Wiese, Nucl.\ Phys.\ B \textbf{729}, 317 (2005).

\bibitem[Bruegger at al. (2006a)]{BKMPW06}
C.\ Br\"ugger, F.\ K\"ampfer, M.\ Moser, M.\ Pepe and U.-J.\ Wiese, Phys.\ Rev.\ B \textbf{74}, 224432
(2006).

\bibitem[Bruegger et al. (2006)]{BKPW06}
C.\ Br\"ugger, F.\ K\"ampfer, M.\ Pepe and U.-J.\ Wiese, Eur.\ Phys.\ J.\ B \textbf{53}, 433 (2006).

\bibitem[Bruegger et al. (2007a)]{BHKPW07}
C.\ Br\"ugger, C.\ P.\ Hofmann, F.\ K\"ampfer, M.\ Pepe and U.-J.\ Wiese, Phys.\ Rev.\ B \textbf{75}, 014421
(2007).

\bibitem[Bruegger et al. (2007b)]{BHKMPW07}
C.\ Br\"ugger, C.\ P.\ Hofmann, F.\ K\"ampfer, M.\ Moser, M.\ Pepe and U.-J.\ Wiese, Phys.\ Rev.\ B
\textbf{75}, 214405 (2007).

\bibitem[Jiang et al. (2009)]{JKHW09}
F.-J.\ Jiang, F.\ K\"ampfer, C.\ P.\ Hofmann and U.-J.\ Wiese, Eur.\ Phys.\ J.\ B \textbf{69}, 473 (2009).

\bibitem[Leutwyler (1997)]{Leu97}
H.\ Leutwyler, Helv.\ Phys.\ Acta \textbf{70}, 275 (1997).

\bibitem[Burgess and Lutken (1998)]{BL98}
C.\ P.\ Burgess, and C.\ A \ Lutken, Phys.\ Rev.\ B \textbf{57}, 8642 (1998).

\bibitem[Son (1994)]{Son94}
D.\ T.\ Son, Phys.\ Rev.\ Lett.\ \textbf{94}, 175301 (2005).

\bibitem[Brauner (2010)]{Brau10}
T.\ Brauner, Symmetry \textbf{2}, 609 (2010).

\bibitem[Burgess (2007)]{Bur07}
C.\ P.\ Burgess, Annu.\ Rev.\ Nucl.\ Part.\ Sci.\ \textbf{57}, 329 (2007); Phys.\ Rept.\ \textbf{330}, 193
(2000).

\bibitem[Goity (2004)]{Goi04}
J.\ L.\ Goity, Czech.\ J.\ Phys. \textbf{51}, B35 (2001).

\bibitem[Scherer (2003)]{Sch03}
S.\ Scherer, Adv.\ Nucl.\ Phys.\ \textbf{27}, 277 (2003).

\bibitem[Manohar (1997)]{Man97}
A.\ V.\ Manohar, in {\it Perturbative and Nonperturbative Aspects of Quantum Field Theory},
edited by H.\ Latal and W.\ Schweiger (Springer, New York, 1997), p.311.

\bibitem[Leutwyler (1995)]{Leu95}
H.\ Leutwyler, in {\it Hadron Physics 94 -- Topics on the Structure and Interaction of Hadronic Systems},
edited by V.\ E.\ Herscovitz, C.\ A.\ Z.\ Vasconcellos and E.\ Ferreira (World Scientific, Singapore, 1995),
p. 1.

\bibitem[Ecker (1995)]{Eck95}
G.\ Ecker, Prog.\ Part.\ Nucl.\ Phys.\ \textbf{35}, 1(1995).

\bibitem[Gerber et al. (2010)]{GHKW10}
U.\ Gerber, C.\ P.\ Hofmann, F.\ K\"ampfer and U.-J.\ Wiese, Phys.\ Rev.\ B \textbf{81}, 064414 (2010).

\bibitem[Wiese and Ying (2011)]{WJ94}
U.-J.\ Wiese and H.\ P.\ Ying, Z.\ Phys.\ B \textbf{93}, 147 (1994).

\bibitem[Gerber et al. (2009)]{GHJNW09}
U.\ Gerber, C.\ P.\ Hofmann, F.-J.\ Jiang, M.\ Nyfeler and U.-J.\ Wiese, J. Stat.\ Mech.: Theory Exp. 2009,
P03021.

\bibitem[Jiang and Wiese (2011)]{JW11}
F.-J.\ Jiang and U.-J.\ Wiese, arXiv:1011.6205.

\bibitem[Gerber et al. (2011)]{GHJPSW11}
U.\ Gerber, C.\ P.\ Hofmann, F.-J.\ Jiang, G.\ Palma, P.\ Stebler and U.-J.\ Wiese, arXiv:1102.3317.

\bibitem[Lange (1966)]{Lan66}
R.\ V.\ Lange, Phys.\ Rev.\ Lett.\ \textbf{14}, 3 (1965); Phys.\ Rev.\ \textbf{146}, 301 (1966).

\bibitem[Guralnik et al. (1968)]{GHK68}
G. S. Guralnik, C. R. Hagen and T. W. B. Kibble, in {\it Advances in Particle Physics}, edited by R. L. Cool
and R. E. Marshak (Wiley, New York, 1968), Vol. 2, p. 567.

\bibitem[Chadha and Nielsen (1976)]{CN76}
H.\ B.\ Nielsen and S.\ Chadha, Nucl.\ Phys.\ B \textbf{105}, 445 (1976).

\bibitem[Schaefer at al. (2001]{Sch01}
T.\ Sch\"afer, D.\ T.\ Son, M.\ A.\ Stephanov, D.\ Toublan and J.\ J.\ M.\ Verbaarschot, Phys.\ Lett.\ B
\textbf{522}, 67 (2001).

\bibitem[Brauner (2007)]{Bra07}
T.\ Brauner, Phys.\ Rev.\ D \textbf{75}, 105014 (2007).

\bibitem[Leutwyler (1994b)]{Leu94b}
H.\ Leutwyler, Ann.\ Phys.\ (N.Y.) \textbf{235}, 165 (1994).

\bibitem[Leutwyler (1989)]{Leu89}
H.\ Leutwyler, Nucl.\ Phys.\ B, Proc.\ Suppl. \textbf{4}, 248 (1988); in {\it New Theories in Physics},
Warsaw International Symposium on Elementary Particle Physics, Kazimierz, 1988, edited by Z.\ Ajduk, S.\
Pokorski and A.\ Trautman (World Scientific, Singapore, 1989), p. 116; also in {\it Symmetry Violations in
Subatomic Physics}, Summer Institute in Theoretical Physics, Kingston, 1988, edited by B.\ Castel and P.\
J.\ O'Donnell (World Scientific, Singapore, 1989), p. 57.

\bibitem[Landsman and van Weert (1987)]{LW87}
N.\ P.\ Landsman and C.\ G.\ van Weert, Phys.\ Rep.\ \textbf{145}, 141 (1987).

\bibitem[Kapusta (1989)]{Kap98}
J.\ I.\ Kapusta, {\it Finite-Temperature Field Theory} (Cambridge University Press, Cambridge, England
1989).

\bibitem[Smilga (1997)]{Smi97}
A.\ V.\ Smilga, Phys.\ Rep.\ \textbf{291}, 1 (1997).

\bibitem[Gerber and Leutwyler (1989)]{GL89}
P.\ Gerber and H.\ Leutwyler, Nucl.\ Phys.\ B \textbf{321}, 387 (1989).

\bibitem[Morita and Tanaka (1965b)]{MT65b}
T.\ Morita and T.\ Tanaka, J.\ Math.\ Phys.\ \textbf{6}, 1152 (1965).

\bibitem[Chang (2001)]{Cha01}
C.\ Chang, Ann.\  Phys.\ \textbf{293}, 111 (2001).

\bibitem[Achleitner (2011)]{Ach11}
J.\ Achleitner, arXiv:1103.1831.

\end{thebibliography}
\end{document}